\documentclass[11pt]{article}

\usepackage{amsmath}
\usepackage{graphicx}
\usepackage{indentfirst}
\usepackage{amssymb}

\usepackage{cite}
\usepackage{color}
\usepackage{subfigure}
\usepackage[colorlinks,linkcolor=blue,citecolor=blue,urlcolor=blue,hyperindex]{hyperref}
\usepackage{indentfirst}
\usepackage{amsmath}
\usepackage{amssymb}
\usepackage{latexsym}
\usepackage{amsmath}
\usepackage{color}
\usepackage{lineno}
\usepackage{graphicx}
\usepackage{float}
\usepackage{subfigure}
\usepackage{geometry}
\usepackage{graphics}
\usepackage{subfigure}
\usepackage{graphicx}
\usepackage{multirow}
\usepackage{booktabs}
\usepackage{hyperref}
\usepackage{cite}
\graphicspath{{AdSBH/},{D=4/},{D=5/},{D=6/},{D=7/}}
\usepackage{array}
\usepackage{caption}
\usepackage{amsthm,amsmath,amssymb}
\usepackage{mathrsfs}
\usepackage{textcomp}
\usepackage{amsmath}
\usepackage{amssymb}

\begin{document}
	
\title{Triple points and phase transitions of $D$-dimensional dyonic AdS black holes with quasitopological electromagnetism in Einstein-Gauss-Bonnet gravity}
	
\date{}
\maketitle

\begin{center}
\author{Ping-Hui Mou}$^{a,}$ $\footnote{\texttt{mph2022@163.com}}$, \author{Qing-Quan Jiang}$^{a,}$ $\footnote{\texttt{Corresponding author:qqjiangphys@yeah.net}}$,
\author{Ke-Jian He}$^{b,}$ $\footnote{\texttt{kjhe94@163.com}}$
and \author{Guo-Ping Li}$^{a,}$ $\footnote{\texttt{Corresponding author:gpliphys@yeah.net}}$
		
\vskip 0.25in
$^{a}$\it{School of Physics and Astronomy, China West Normal University, Nanchong 637000, People's Republic of China}\\
$^{b}$\it{College of Physics, Chongqing University, Chongqing 401331, People's Republic of China}

\end{center}
\vskip 0.6in
{\abstract
	{By considering the negative cosmological constant $\Lambda$ as a thermodynamic pressure, we study the thermodynamics and phase transitions of the $D$-dimensional dyonic AdS black holes (BHs) with quasitopological electromagnetism in Einstein-Gauss-Bonnet (EGB) gravity. 	
	The results indicate that the small/large BH phase transition that is similar to the van der Waals (vdW) liquid/gas phase transition always exist for any space-time dimensions.
	But interestingly, we then find that this BH system exhibits a more complex phase structure in 6-dimensional case that missing in other dimensions. 
	Specifically, it shows for $D=6$ that we observed the small/intermediate/large BH phase transitions in a specific parameter region with the triple point naturally appeared.
	However, for the dyonic AdS BHs with quasitopological electromagnetism in Einstein gravity, the novel phase structure composed of two separate coexistence curves observed in\cite{Li:2022vcd} disappeared in EGB gravity.  
	In addition, it is also true that the critical exponents calculated near the critical points possess identical values as mean field theory.
	Finally, we conclude that these findings shall provide some deep insights into the intriguing thermodynamic properties of the dyonic AdS BHs with quasitopological electromagnetism in EGB gravity.
		 
	}
}

\thispagestyle{empty}
\newpage
\setcounter{page}{1}

\section{Introduction}\label{sec1}
The BH, known for its strong gravity, has been regarded as a perfect object for testing general relativity. 
In recent years, the Laser Interferometer Gravitational
Wave Observatory (LIGO) detected gravitational waves from binary BH mergers\cite{LIGOScientific:2016kms,LIGOScientific:2016sjg}, and the Event Horizon
Telescope (EHT) collaboration released images of supermassive BHs at the centers of galaxy $M87^*$ and the Milky Way\cite{EventHorizonTelescope:2019dse,EventHorizonTelescope:2019uob,EventHorizonTelescope:2019jan,EventHorizonTelescope:2019ths,EventHorizonTelescope:2022wkp,EventHorizonTelescope:2022apq,EventHorizonTelescope:2022wok}. 
Since then, a great deal of interesting research has been attracted to work on gravitational waves\cite{Liu:2020mru,Cai:2021uup,Cai:2023ywp,Xu:2023wog} and BH shadows\cite{Gralla:2019xty,Gralla:2020srx,Li:2021riw,Gan:2021xdl,Zeng:2021mok,Hadar:2022xag,Wang:2022yvi,Hou:2022eev,Chen:2022scf}. These studies does not only obtain some important theoretical results, but also strongly demonstrate the existence of BHs in our universe and successfully validate the predictions of general relativity. 
In fact, it had been previously demonstrated that BH is a thermodynamic system, and it possess numerous interesting and important thermodynamic properties that the ordinary thermodynamic system does not have\cite{Bardeen:1973gs,Bekenstein:1973ur,Hawking:1975vcx}. In particular, Hawking and Page in the early 1980s showed that there is a phase transition between a stable BH and thermal radiation in the Schwarzschild-AdS BH, which is known as the Hawking-Page phase transition\cite{Hawking:1982dh}. The anti-de Sitter/conformal field theory (AdS/CFT) correspondence suggests that the thermodynamic of the BH in AdS space can be related to the thermodynamic of the dual strongly coupled conformal field theory in the boundary of AdS space\cite{Maldacena:1997re,Gubser:1998bc,Witten:1998qj}. Therefore, within the framework of AdS/CFT correspondence, Witten interpreted the Hawking-Page phase transition as a confinement/deconfinement phase transition of gauge field\cite{Witten:1998zw}. So, the phase transition, as an interesting property of BH has been widely studied for a long time\cite{Chamblin:1999tk,Chamblin:1999hg,Surya:2001vj,Cho:2002hq,Shen:2007xk,Cai:2007wz}.

Recently, the thermodynamics of AdS BHs have been studied in the extended phase space. In this framework, the negative cosmological constant $\Lambda$ is treated as the thermodynamic pressure of BH, with its conjugate quantity being the thermodynamic volume of BH\cite{Kastor:2009wy,Dolan:2011xt,Cvetic:2010jb,Dolan:2013ft,Castro:2013pqa,Kastor:2010gq,MahmoudEl-Menoufi:2013lfe}. Subsequently, by comparing with the liquid–gas system, Kubiz\v{n}\'{a}k and Mann proved that the BH system have the similar features, such as the same oscillatory behavior of pressure-volume ($P-V$), critical exponents and scaling relations\cite{Kubiznak:2012wp}. Therefore, an accurate analogy has been established between the charged AdS BH and the vdW system\cite{Kubiznak:2012wp}. 
This analogy has also been generalized to different types of AdS BHs, and then the later results show that the small/large BH phase transition similar to the vdW liquid/gas phase transition are generally obtained in various AdS BHs\cite{Gunasekaran:2012dq,Chen:2013ce,Cai:2013qga,Mo:2014mba,Zou:2014mha,Wei:2015iwa}.
Afterwards, a series of interesting phase transitions were discovered in the extended phase space, such as the reentrant phase transition and the triple point\cite{Altamirano:2013ane,Altamirano:2013uqa,Wei:2014hba,Frassino:2014pha,Zhang:2017lhl,Dehyadegari:2017hvd,Dehghani:2020blz,Zhang:2020obn,Cui:2021qpu,Wei:2021krr,Li:2022vcd,Liu:2022spy,Qu:2022nrt,Bai:2022vmx}. And in the case of higher-order gravity theories, there are also multicritical phase transitions in AdS BHs, i.e., in higher-order Lovelock gravity\cite{Tavakoli:2022kmo,Wu:2022bdk,Wu:2022plw}. 

The extended gravity theories, including EGB gravity, have brought some new insights to study the property of BH\cite{Cai:2013qga,Zou:2014mha,Wei:2014hba,Mo:2014mba,Zhang:2020obn,Lin:2020kqe,Yang:2020jno,Ghosh:2016ddh,EslamPanah:2020hoj,Fernandes:2020rpa,Glavan:2019inb,Yerra:2022eov,Singh:2021xbk,Singh:2022dth}. 
As a higher-order extension of Einstein gravity, the EGB gravity exhibits several striking features.
In Ref. \cite{Glavan:2019inb}, Glavan introduced a rescaled GB coupling constant and defined the four-dimensional theory as a limit of $D\rightarrow4$, thereby extended the Einstein-Hilbert action by incorporating the GB term. This framework bypasses the Lovelock theory and avoids Ostrogradsky instability\cite{Glavan:2019inb}.  
Recently, some significant progresses have been made in the research of BHs in the background of EGB gravity\cite{Zhang:2020obn,Lin:2020kqe,Yang:2020jno,Ghosh:2016ddh,EslamPanah:2020hoj,Fernandes:2020rpa,Yerra:2022eov,Singh:2021xbk,Singh:2022dth}. For instance, in four-dimensional EGB gravity, there exist triple points in the Born-Infeld AdS BH\cite{Zhang:2020obn}. These studies indicate that BHs in EGB gravity possess many interesting properties. 
On the other hand, the quasitopological electromagnetism is also regarded as an important and interesting object in the study of BHs\cite{Liu:2019rib,Cisterna:2020rkc,Lei:2020clg,Cisterna:2021ckn,Barrientos:2022uit}.
Recently, Liu et al. proposed a new concept of quasitopological electromagnetism, and defined it as the square norm of the topological wedge product of Maxwell field strength of order $k$ ($k\geqslant2$)\cite{Liu:2019rib}. 
In general, quasitopological terms have no effects on either the Maxwell equations or the energy-momentum tensor. Although they are known to have no effect on pure electric or magnetic Reissner-Nordstr\"om (RN) BHs, their influence on the dyonic BHs cannot be ignored.
Later, Li et al. investigated the thermodynamics and phase transitions of the dyonic AdS BHs with quasitopological electromagnetism in Einstein gravity\cite{Li:2022vcd}. The result shows that the triple points can be observed, as well as a novel phase structure composed of two separate coexistence curves. Obviously, quasitopological electromagnetism has a significant effect on the phase transitions of dyonic BHs.

Very recently, combining the quasitopological electromagnetism with EGB gravity, a solution of $D$-dimensional dyonic AdS BHs has 
been obtained\cite{Sekhmani:2022lws}. Interestingly, the authors found that it is thermodynamically stable in certain regions, while unstable in others. 
Also, the shadow of this BH showed that the radius of the BH shadow decreased with the  GB coupling constant $\alpha$ or the dimension $D$. 
Moreover, it shows that the GB coupling constant $\alpha$ and magnetic charge $Q_{m}$ have some significant effects on energy emission rates, while other quantities have relatively small effects. Clearly, these studies means that the dyonic AdS BHs with quasitopological electromagnetism in EGB gravity have many intriguing characteristics, which is worth investigating it more deeply.
In particular, the thermodynamic properties and phase transitions of the dyonic AdS BHs with quasitopological electromagnetism in EGB gravity still unknown. And, this BH may exhibit some rich phase transition behaviors due to the dimensional $D$ and the corresponding coupling constants.
Therefore, this paper aims to study the thermodynamics and phase transitions of the dyonic AdS BHs in the extended phase space. We hope that through our research, we can further reveal the thermodynamic characteristics of this BH and then further provide some new insights into understanding the effects of quasitopological electromagnetism on dyonic AdS BHs.

This paper is organized as follows. In Sec.\ref{sec2}, we review the $D$-dimensional dyonic AdS BHs with quasitopological electromagnetism in EGB gravity and study its thermodynamics. In Sec.\ref{sec3}, we investigate the phase transitions and phase diagrams of the $D$-dimensional dyonic AdS BHs. In Sec.\ref{sec4}, we calculate the critical exponents near the critical points. Sec.\ref{sec5} concludes with a summary and discussion.

\section{Thermodynamics of the $D$-dimensional dyonic AdS BHs}\label{sec2}
The action of $D$-dimensional EGB gravity minimally coupled to the quasitopological electromagnetism can be expressed as\cite{Sekhmani:2022lws}
\begin{align}
	S_{D}=\frac{1}{16\pi}\int d^{D}x\sqrt{-g}(\mathcal{R}-2\Lambda+\alpha \mathcal{G}+\mathcal{L}_{\rm{QE}}),
\end{align}
where, $\Lambda=-\frac{(D-1)(D-2)}{2l^{2}}$ is the cosmological constant in the AdS space, $\mathcal{G}=\mathcal{R}^{2}-4\mathcal{R}^{\mu\nu}\mathcal{R}_{\mu\nu}+\mathcal{R}^{\mu\nu\rho\sigma}\mathcal{R}_{\mu\nu\rho\sigma}$ is the GB term, $\alpha$ is the GB coupling constant with dimension [length]$^{2}$. And, the matter field is represented by the following Quasitopological Electromagnetism Lagrangian,
\begin{align}
\mathcal{L}_{\rm{QE}}=-(\frac{1}{4}\mathcal{F}^{2}+\frac{1}{2p!}\mathcal{H}^{2}+\beta\mathcal{L}_{\rm{int}})
\end{align}
with $p=D-2$, $\mathcal{F}^{2}=\mathcal{F}_{\mu\nu}\mathcal{F}^{\mu\nu}$ and $\mathcal{H}^{2}=\mathcal{H}_{\rho_{1}\ldots\rho _{p}}\mathcal{H}^{\rho_{1}\ldots\rho _{p}}$, and $\beta$ is the coupling constant of the dimension [length]$^{2}$. And, the interaction term is given by
\begin{align}
	\mathcal{L}_{\rm{int}}=\delta_{\gamma_{1}\ldots\gamma _{D}}^{\lambda_{1}\ldots\lambda _{D}}\mathcal{F}_{\lambda_{1}\lambda_{2}}\mathcal{H}_{\lambda_{3}\ldots\lambda _{D}}\mathcal{F}^{\gamma_{1}\gamma_{2}}\mathcal{H}^{\gamma_{3}\ldots\gamma _{D}}. 
\end{align}
The above action admits a BH solution with the form expressed as the following metric,
\begin{align}
	ds^2=-f(r)dt^2+f(r)^{-1}dr^2 +r^2 d\Omega_{D-2} ^2 \ 
\end{align}
where $d\Omega_{D-2}^{2}$ is the ($D-2$)-dimensional unit sphere. The metric function is given by
\begin{align}
f(r)=1+\frac{r^2}{2\widetilde{\alpha}} \left (1- \sqrt{1+\frac{4\widetilde{\alpha}m}{r^{D-1}}+\frac{8\widetilde{\alpha}\Lambda}{(D-2)(D-1)}-\frac{2 \widetilde{\alpha}}{D-3}\left( Q_{m}^{2}+Q_{e}^{2} \mathcal{E}\right) }\right ),
\end{align}
where
\begin{align}
\mathcal{E}=\ _2F_1\left[1,\frac{D-3}{2(D-2)};\frac{7-3D}{4-2D};\frac{-8\beta Q_{m}^{2}\Gamma(D-1)^{2}}{r^{2D-4}}\right]
\end{align}
is the hypergeometric function.
Here, $\widetilde{\alpha}=(D-4)(D-3)\alpha$, $Q_{e}$ and $Q_{m}$ are the BH electric charge and magnetic charge, respectively. The parameter $m$ is interpreted as the mass of the solution in a specific parameter space, which is the integration constant provided by the boundary conditions. Thus, the relevant Arnowitt-Deser-Misner (ADM) mass $M$ is defined as\cite{Ghosh:2014pga}
\begin{align}
	M=\frac{(D-2)\mathcal{V}_{D-2}}{16\pi}m,
\end{align}
where $\mathcal{V}_{D-2}=\frac{2\pi^{\frac{D-1}{2}}}{\Gamma\left( \frac{D-1}{2}\right) }$ is the volume of the ($D-2$)-dimensional unit
sphere.
In the extended phase space, the negative cosmological constant $\Lambda$ is treated as the thermodynamic pressure $P=-\frac{\Lambda}{8\pi}$\cite{Kastor:2009wy}.
The BH horizon radius $r_{h}$ is determined by the largest root of the equation $f(r_{h})=0$. And, it can be obtained by solving this equation. Therefore, we can express the mass of BH in term of the horizon radius $r_{h}$ as
\begin{align}
	M= \frac{P\pi^{\frac{D-1}{2}}r_{h}^{D-1}}{\Gamma(\frac{D+1}{2})}+ \frac{(D-2)\pi^{\frac{D-3}{2}}r_{h}^{D-3}}{8\Gamma(\frac{D-1}{2})}+\frac{\pi^{\frac{D-3}{2}}r_{h}^{3-D}(Q_{m}^{2}+Q_{e}^{2}\mathcal{E}_{r_h})}{16(D-3)\Gamma \left( \frac{D-1}{2}\right) } +\frac{\pi^{\frac{D-3}{2}}r_{h}^{D-5}\mathcal{N}_{1}}{8\Gamma(\frac{D-1}{2})} ,\label{mass}
\end{align}
where
\begin{align}
	\mathcal{E}_{r_{h}}=\ _2F_1\left[1,\frac{D-3}{2(D-2)};\frac{7-3D}{4-2D};\frac{-8\beta Q_{m}^{2}\Gamma(D-1)^{2}}{r_{h}^{2D-4}}\right],\ \mathcal{N}_{1}=(D-4)(D-3)(D-2)\alpha.
\end{align}
Further, based on the definition of Hawking temperature $T=\frac{f'(r_{h})}{4\pi}$, the BH temperature can be obtained as
\begin{align}
	T=\frac{\left[ 12+2D(D-5)+32P\pi r_{h}^{2}+2D(D-5)r_{h}^{-2}\mathcal{N}_{1}-r_{h}^{6-2D}Q_{m}^{2}\right] \mathcal{N}_{2}-r_{h}^{6}Q_{e}^{2}}{\left[ 8(D-2)\pi r_{h}+16\pi r_{h}^{-1}\mathcal{N}_{1}\right] \mathcal{N}_{2}},
	\label{temperature}
\end{align}
where
\begin{align}
	\mathcal{N}_{2}=r_{h}^{2D}+8r_{h}^4\beta Q_{m}^2\Gamma(D-1)^2.
\end{align}
It is worth note that BH mass in the extended phase space should be regarded as the enthalpy, instead of the internal energy of the system, i.e., $H\equiv M$.
Therefore, the other thermodynamic quantities of BH, such as the thermodynamic volume $V$, entropy $S$, electric potential $\Phi_{e}$, and magnetic potential $\Phi_{m}$, can be calculated as follows,
\begin{align}
	V=\left( \frac{\partial H}{\partial P}\right) _{S,Q_{e},Q_{m}}=\frac{2\pi^{\frac{D-1}{2}}r_{h}^{D-1}}{(D-1)\Gamma\left( \frac{D-1}{2}\right) },
\end{align}
\begin{align}
	S=\int_{0}^{r_{h}}T^{-1}\left( \frac{\partial H}{\partial r}\right) _{P,Q_{e},Q_{m}}dr=\frac{\pi^{\frac{D-1}{2}} r_{h}^{D-2}\left[1+2(D-3)(D-2)\alpha r_{h}^{-2} \right] }{2\Gamma \left( \frac{D-1}{2}\right) },\label{entropy}
\end{align}
\begin{align}
	\Phi_{e}=\left( \frac{\partial H}{\partial Q_{e}}\right) _{S,P,Q_{m}}=\frac{\pi^{\frac{D-3}{2}}r_{h}^{3-D}Q_{e} \mathcal{E}_{r_h}  }{8(D-3)\Gamma\left( \frac{D-1}{2}\right) },
\end{align}
\begin{align}
\Phi_{m}=\left( \frac{\partial H}{\partial Q_{m}}\right) _{S,P,Q_{e}}=\frac{\pi^{\frac{D-3}{2}}r_{h}^{3-D} \left[   2(D-2)Q_{m}^{2}+(D-3)Q_{e}^{2} \left(  r_{h}^{2D}\mathcal{N}_{2}^{-1}- \mathcal{E}_{r_h}\right)  \right]   }{16(D-3)(D-2)Q_{m}\Gamma\left( \frac{D-1}{2}\right) }.
\end{align}
By utilizing these thermodynamic quantities, it is easy to verify the validity of the first law of thermodynamics, which takes the form of 
\begin{align}
	dH=TdS+\Phi_{e}dQ_{e}+\Phi_{m}dQ_{m}+ \Phi_{\alpha}d\alpha+\Phi_{\beta}d\beta+VdP,
\end{align}
where
\begin{align}
	\Phi_{\alpha}\equiv \left(  \frac{\partial H}{\partial \alpha}\right)  _{S,P,Q_{e},Q_{m},\beta}=\frac{\pi^{\frac{D-3}{2}}r_{h}^{D-5}\mathcal{N}_{1}}{8\alpha\Gamma\left( \frac{D-1}{2}\right) }-\frac{(D-3)(D-2)\pi^{\frac{D-1}{2}}r_{h}^{D-4}T}{\Gamma\left( \frac{D-1}{2}\right) }
\end{align}
is the conjugate quantity to the GB coupling constant $\alpha$, and
\begin{align}
	\Phi_{\beta}\equiv\left(  \frac{\partial H}{\partial \beta}\right)  _{S,P,Q_{e},Q_{m},\alpha}=\frac{\pi^{\frac{D-3}{2}}r_{h}^{3-D}Q_{e}^{2}\left(  r_{h}^{2D}\mathcal{N}_{2}^{-1}- \mathcal{E}_{r_h}\right)}{32(D-2)\beta\Gamma\left( \frac{D-1}{2}\right) }
\end{align}
is the conjugate quantity to the coupling constant $\beta$. In this case, $\alpha$ and $\beta$ should be considered as novel thermodynamic variables by considering the dimensional characteristic of them. Moreover, the generalized Smarr relation can be derived as
\begin{align}
		(D-3)H=(D-2)TS-2PV+(D-3)\Phi_{e}Q_{e}+(D-3)\Phi_{m}Q_{m}+2\alpha\Phi_{\alpha} +2\beta\Phi_{\beta}.
\end{align}
The Gibbs free energy, which is a quantity measures the global stability of BH system, is give as $G= H-TS$, which is, 

\begin{align}
	G&=2(D-2)r_{h}^{2D+2}\mathcal{N}_{3}+2r_{h}^{2D}\mathcal{N}_{1}\mathcal{N}_{3}+\frac{32P\pi r_{h}^{2D+4}\mathcal{N}_{3}}{D-1}+\frac{r_{h}^{8}(Q_{m}^{2}+Q_{e}^2\mathcal{E}_{r_h})\mathcal{N}_{3}}{D-3}\nonumber \\
	&\quad -\frac{\left[ 2(D-3)(D-2)\alpha+r_{h}^{2}\right] \left[ (2r_{h}^{2D}\mathcal{N}_{4}-r_{h}^{8}Q_{m}^{2})\mathcal{N}_{2}-r_{h}^{2D+8}Q_{e}^{2}\right] \mathcal{N}_{3}}{\left[ (D-2)r_{h}^{2}+2\mathcal{N}_{1}\right] \mathcal{N}_{2}} ,
\end{align}
where
\begin{align}
	\mathcal{N}_{3}=\frac{\pi^{\frac{D-3}{2}}r_{h}^{-5-D}}{16\Gamma\left( \frac{D-1}{2}\right) },\
	\mathcal{N}_{4}=6r_{h}^2+D(D-5)r_{h}^2+16P\pi r_{h}^4+(D-5)\mathcal{N}_{1}.
\end{align}
As is well known, the swallowtail behavior of the Gibbs free energy indicates the occurrence of a first-order phase transition in BH systems. And multiple swallowtails mean some rich varieties of phase transitions in the system, including the triple points. Therefore, in this paper, we study the phase transitions of BHs by analyzing the swallowtail behavior of the Gibbs free energy.

\section{Phase transitions and phase diagrams of the $D$-dimensional dyonic AdS BHs }\label{sec3}
In this section, we would like to study the phase transitions and phase diagrams of the dyonic AdS BHs with quasitopological electromagnetism in dimensions $D=4$, $5$, $6$ and $7$.
Based on the equation of temperature (\ref{temperature}), the equation of state for BH system can be obtained as 
\begin{align}
	P&= \frac{Q_{m}^{2}}{32\pi r_{h}^{2D-4}}-\frac{(D-4)(D-3)(D-2)(D-5-8\pi r_{h}T)\alpha}{16\pi r_{h}^{4}}
	 \nonumber \\
	&\quad+\frac{(D-2)(3-D+4\pi r_{h}T)}{16\pi r_{h}^{2}}+\frac{Q_{e}^{2}}{32\pi r_{h}^{2D-4}+256\pi \beta Q_{m}^{2} \Gamma(D-1)^{2}} .\label{pressure}
\end{align}
According to \cite{Kubiznak:2012wp}, the specific volume can be given as $v=\frac{4r_{h}}{D-2}$, thereby the equation of state can be rewritten in the form, 
\begin{align}
	P(v,T)&= \frac{16(D-4)(D-3)\left[5-D+ 2(D-2)\pi vT \right] \alpha}{(D-2)^{3}\pi v^{4}}+\frac{2^{4D-13}(D-2)^{4-2D}v^{4-2D}Q_{m}^{2}}{\pi}
	\nonumber \\
	&\quad +\frac{3-D+(D-2)\pi vT}{(D-2)\pi v^{2}} +\frac{Q_{e}^{2}}{2^{13-4D}(D-2)^{2D-4}\pi v^{2D-4}+256\pi\beta Q_{m}^{2}\Gamma(D-1)^{2}}.
\end{align}
Evidently, there exists a direct proportionality between the specific volume $v$ and horizon radius $r_{h}$. Therefore, we will utilize the equation of state (\ref{pressure}) in the following calculations.
In addition, since the thermodynamic volume $V\varpropto r_{h}^{D-1}$, the critical point can be determined by the conditions,
\begin{align}
	\left(  \frac{\partial P}{\partial r_{h}}\right) _{T} =0  ,\
	\left(  \frac{\partial^2P}{\partial r_{h}^2}\right) _{T} =0; \ or\	\left(  \frac{\partial T}{\partial r_{h}}\right) _{P} =0  ,\
	\left(  \frac{\partial^2T}{\partial r_{h}^2}\right) _{P} =0. \label{conditions}
\end{align}	
Different from the Gibbs free energy $G$, the heat capacity  with a constant pressure $C_{P}$, is a quantity used to describe the local thermodynamic stability of BHs. Specifically, a positive value of $C_{P}$ indicates that the system is locally stable, while a negative value is local instability.
At a fixed pressure $P$, the heat capacity $C_{P}$ is defined as
 \begin{align}
 	C_{P}=T\left( \frac{\partial S}{\partial T}\right) _{P}.
 \end{align}
Further, $C_{P}$ can be also reexpressed as 
\begin{align}
	C_{P}=T\left(  \frac{\partial_{r_{h}}S}{\partial_{r_{h}}T}\right)  _{P} \varpropto (\partial_{r_{h}}T)_{P}^{-1}.
\end{align}
Here, we only considered the range of $T>0$ and $S>0$ in this paper. Therefore, it can be see that the positive (negative) slopes of the BH branches on the $T-r_{h}$ plane correspond to the thermodynamically stable (unstable) phases, wherein their corresponding heat capacity $C_{P}$ being positive (negative).
In the following, our attention will mainly focus on the phase transitions and phase diagrams.
For convenience, we will take the case, i.e., the magnetic charge $Q_{m}=2$ and electric charge $Q_{e}=1$, as an example to further study the phase transitions and phase diagrams of the dyonic AdS BHs in this paper. Specifically, we will discuss the phase transitions and phase diagrams for different dimensions and different values of the coupling parameters ($\alpha$ and $\beta$).

\subsection{For the $4$-dimensional case}\label{sec31}
\begin{figure} [H]
	\centering
	\subfigure[]{
		\includegraphics[width=0.28\textwidth]{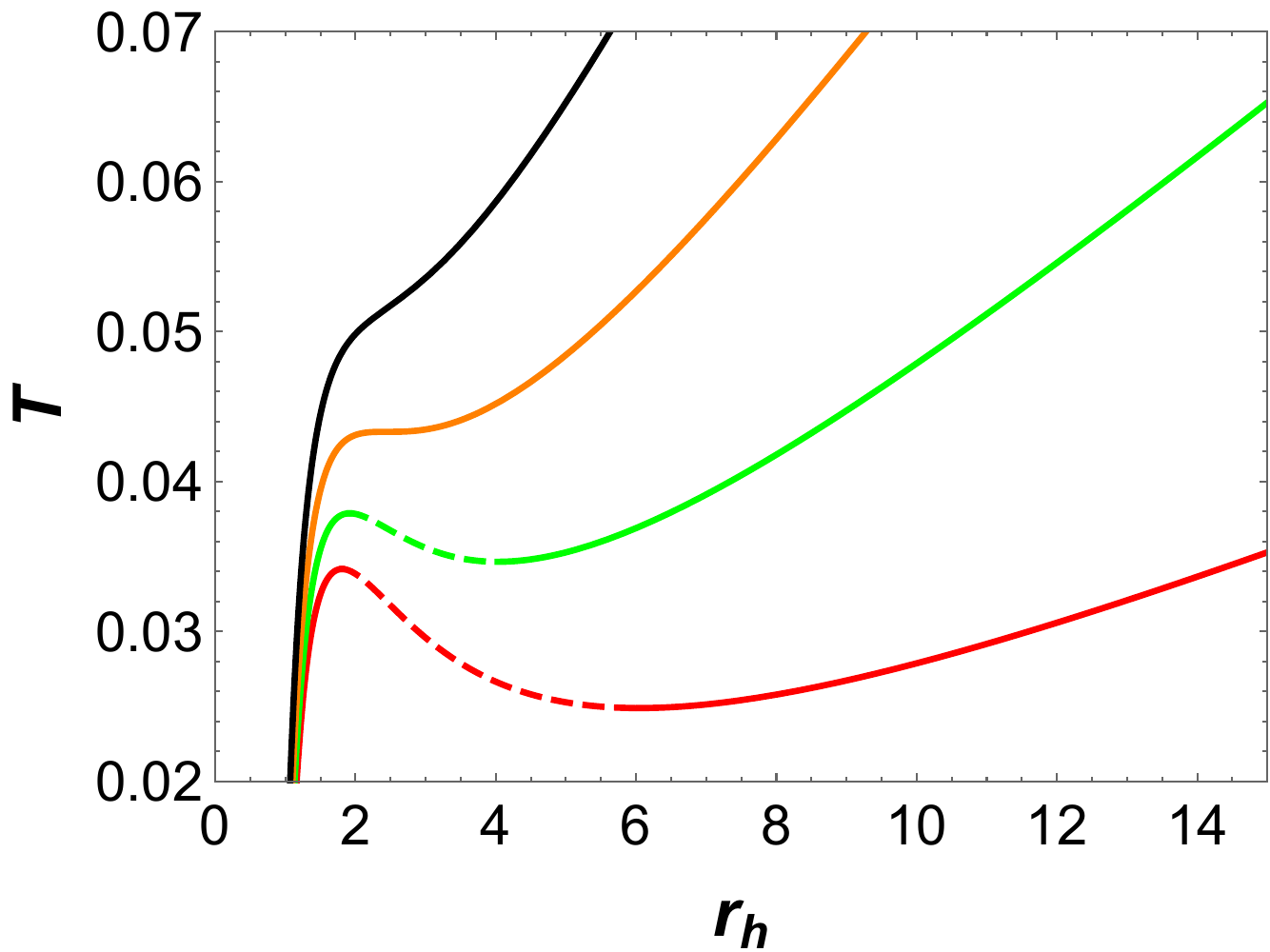}}
	\subfigure[]{
		\includegraphics[width=0.28\textwidth]{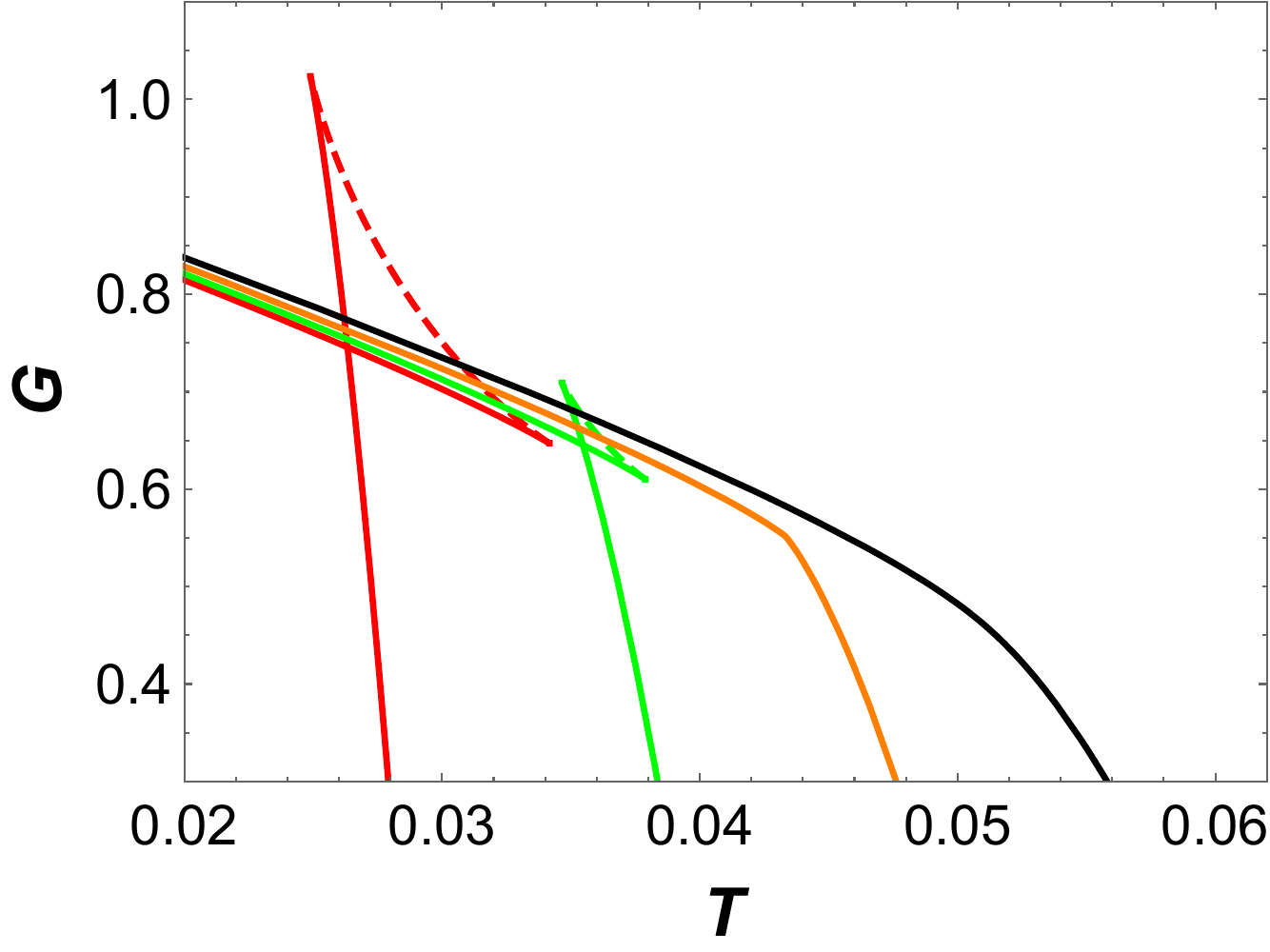}}
	\subfigure[]{
		\includegraphics[width=0.28\textwidth]{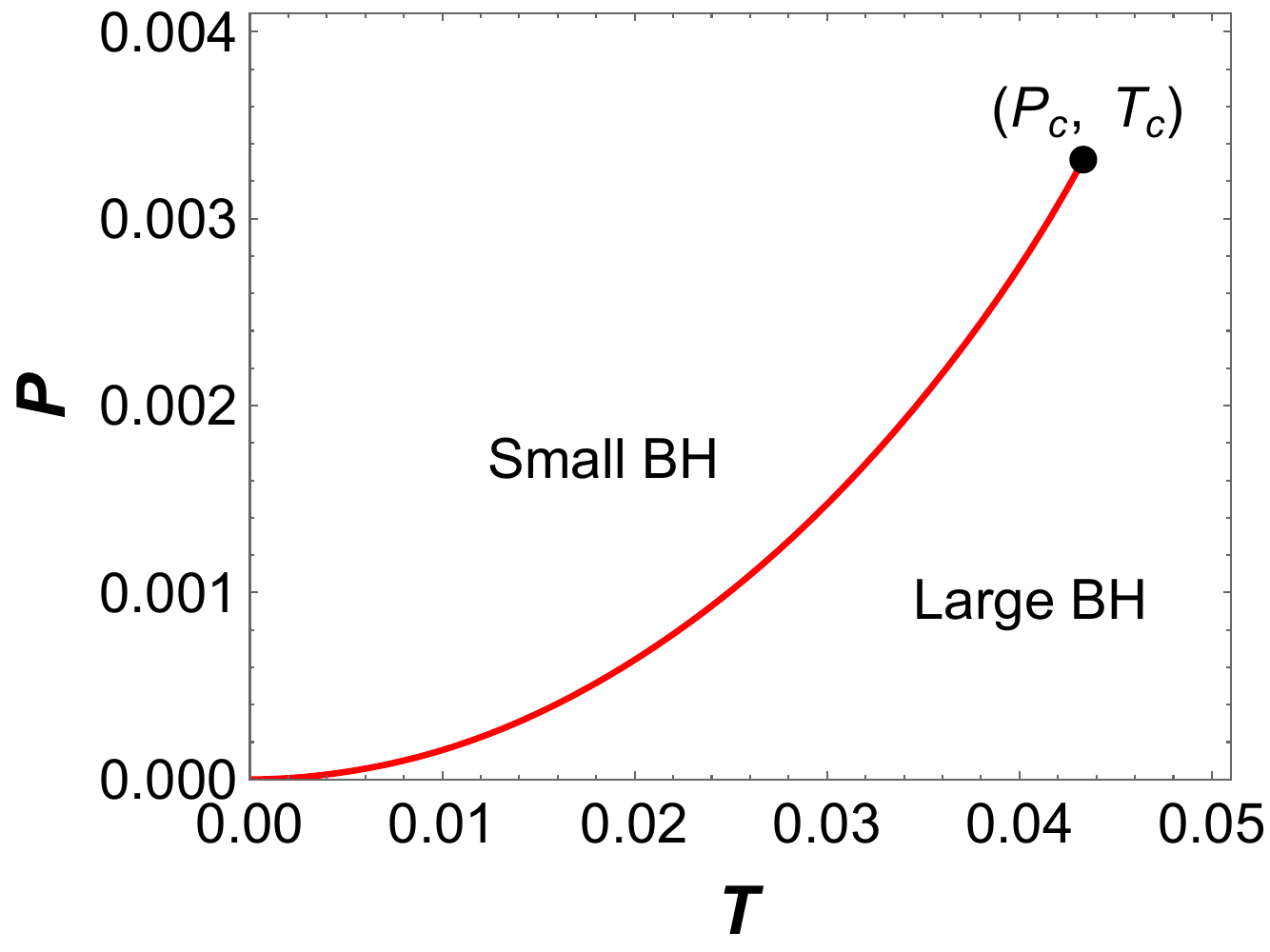}}
	\caption{\small (a) Temperature $T$ vs. $r_{h}$. (b) Gibbs free energy $G$ vs. $T$. The red, green, orange and black isobaric curves in (a) and (b) correspond to $P=0.001$, $0.002$, $0.0033158$ and $0.005$, respectively. Solid curves indicate stable branches, while dashed curves represent unstable branches. (c) Phase diagram for the 4-dimensional dyonic AdS BH when $\alpha=0.5$ and $\beta=0.1$.}\label{fig1}
\end{figure}

In this subsection, we would like to study the phase transitions and phase diagrams of the 4-dimensional dyonic AdS BH. According to Eqs.(\ref{pressure}) and (\ref{conditions}), we obtained the critical points for different values of GB coupling constant $\alpha$ and $\beta$ as shown in Tab.\ref{ta1}.
\begin{table}[H]
	\centering
	\captionsetup{font=footnotesize}
	{\caption {Critical points for different values of coupling constants $\alpha$ and $\beta$, when the dimension $D=4$.}
		{\footnotesize	\vspace{1mm}
			
			\begin{tabular} {ccc|ccc}
				\hline 
				{$\alpha$}  &{$\beta$}     &{$(T_{c},P_{c})$}  &{$\alpha$}  &{$\beta$}     &{$(T_{c},P_{c})$}   \\ 
				\hline
				{0.01}  &{0.01}  &{(0.04331644704, 0.003316488235)}  &{0.1}  &{0.01}   &{(0.04331644704, 0.003316488235)}  \\
				{0.01}  &{0.05}  &{(0.04331648727, 0.003315881743)}  &{0.1}  &{0.05}   &{(0.04331648727, 0.003315881743)}  \\
				{0.01}  &{0.1}   &{(0.04331648854, 0.003315804969)}  &{0.1}  &{0.1}    &{(0.04331648854, 0.003315804969)}   \\
				{0.01}  &{0.5}   &{(0.04331648894, 0.003315743396)}  &{0.1}  &{0.5}    &{(0.04331648894, 0.003315743396)}   \\
				{0.01}  &{0.9}   &{(0.04331648895, 0.003315736546)}  &{0.1}  &{0.9}    &{(0.04331648895, 0.003315736546)}   \\
				{0.01}  &{1}     &{/}                                &{0.1}  &{1}      &{/}                                 \\
				\hline
				{0.5}	&{0.01}  &{(0.04331644704, 0.003316488235)}  &{1}    &{0.01}   &{(0.04331644704, 0.003316488235)}  \\
				{0.5}	&{0.05}  &{(0.04331648727, 0.003315881743)}  &{1}    &{0.05}   &{(0.04331648727, 0.003315881743)}  \\
				{0.5}	&{0.1}   &{(0.04331648854, 0.003315804969)}  &{1}    &{0.1}    &{(0.04331648854, 0.003315804969)}  \\
				{0.5}	&{0.5}   &{(0.04331648894, 0.003315743396)}  &{1}    &{0.5}    &{(0.04331648894, 0.003315743396)}  \\
				{0.5}	&{0.9}   &{(0.04331648895, 0.003315736546)}  &{1}    &{0.9}    &{(0.04331648895, 0.003315736546)}   \\
				{0.5}   &{1}     &{/}                                &{1}    &{1}      &{/}                                 \\
				\hline
				{10}	&{0.01}  &{(0.04331644704, 0.003316488235)}  &{50}   &{0.01}   &{(0.04331644704, 0.003316488235)}  \\
				{10}	&{0.05}  &{(0.04331648727, 0.003315881743)}  &{50}   &{0.05}   &{(0.04331648727, 0.003315881743)}  \\
				{10}	&{0.1}   &{(0.04331648854, 0.003315804969)}  &{50}   &{0.1}    &{(0.04331648854, 0.003315804969)}  \\
				{10}	&{0.5}   &{(0.04331648894, 0.003315743396)}  &{50}   &{0.5}    &{(0.04331648894, 0.003315743396)}  \\
				{10}	&{0.9}   &{(0.04331648895, 0.003315736546)}  &{50}   &{0.9}    &{(0.04331648895, 0.003315736546)}   \\
				{10}    &{1}     &{/}                                &{50}   &{1}      &{/}                                  \\
				\hline 
			\end{tabular}\label{ta1}}		
}\end{table}
From Tab.\ref{ta1}, one can see that each situation has either one critical point or no critical point. 
Specifically, there is no critical point present in the region as long as $\beta\geqslant1$, whatever the value of $\alpha$ is. 
The critical points changed slightly with the $\beta$, while is hardly changed with $\alpha$. 
Since there is only one critical point for different values of parameters $\alpha$ and $\beta$, we can predicate that the BH system only exhibits single phase structure in 4-dimensional case.
For convenience, we take $\alpha=0.5$ and $\beta=0.1$ as an example to study the phase transitions of BHs, where the critical point reads
\begin{align}
	T_{c}=0.0433165,P_{c}=0.0033158.
\end{align}
To analyze the behaviors of temperature $T$ with respect to $r_{h}$ and Gibbs free energy $G$ with respect to $T$, we plotted Figs.\ref{fig1} (a) and (b) by considering some acceptable of $P$, i.e., $P = 0.001$, $0.002$, $0.0033158$ and $0.005$.

Firstly, when $P<P_{c}$, it shows from Fig.\ref{fig1} (a) that two extremal points can be observed on both the red and green isobaric curves, which divide them into three branches: small BH branch, intermediate BH branch, and large BH branch. Both small and large BH branches exhibit positive slope, implies that their heat capacity $C_P$ is positive, and hence, they are thermodynamically stable, which is represented by solid curves in the figure. In contrast, the intermediate BH branch shows a negative slope, suggests that its heat capacity $C_P$ is negative, thereby it is thermodynamic instability, which is represented by a dashed curve. Increasing the pressure to $P>P_{c}$, we observe that there is no extremal point for the black curve. This implies that the system has only one stable BH branch in this case.

Next, we will focus on the behaviors of Gibbs free energy presented in Fig.\ref{fig1} (b). It should be noted that Figs.\ref{fig1} (a) and (b) represent the same BH branchs, where the same color curves correspond to the same pressures. When $P<P_{c}$, both the red and green isobaric curves exhibit the swallowtail behavior, which indicates a first-order small/large BH phase transition occurs. It is worth emphasizing that the nonsmooth points on the isobaric curves in the $G-T$ diagram correspond to the extremal points on the isobaric curves in the $T-r_{h}$ diagram, where dashed curves represent the negative heat capacity and solid curves denote the positive one. For the red and green isobaric curves, the system initially exhibits a small BH phase, and then gradually transforms into a large BH phase as the temperature $T$ increases and passes through the swallowtail intersection. By comparing the red, green, and orange isobaric curves, we find that the swallowtail gradually becomes small and eventually disappears as the pressure $P$ increased. With a progressive increase of pressure $P$, it shows that the extremal points of the temperature on the isobaric curve gradually are close to each other and finally overlapped at the critical pressure. 
When the pressure reaches the critical pressure $P_{c}$, the isobaric curve no longer exhibits swallowtail behavior. When $P>P_{c}$, the Gibbs free energy becomes a monotonic function of temperature, indicating that the system will not undergo any phase transitions.
Finally, the coexistence curve of the small and large BH phases is plotted as Fig.\ref{fig1} (c).
The $P-T$ diagram indicates that the area above the coexistence curve corresponds to the small BH phase, while the area below it corresponds to the large BH phase. It is obvious that the pressure monotonically increased with the temperature, and finally stopped at the critical point $(P_{c}, T_{c})$ at which a second-order phase transition of BH occurred. This small/large BH phase transition is a typical case, which is similar to the vdW liquid/gas phase transition. 
And for other choice of $\alpha$ and $\beta$ in Tab.\ref{ta1}, we have also studied the corresponding phase transitions in 4-dimensional case, and find that these phase transitions always are similar to the case $\alpha=0.5$ and $\beta=0.1$.

\subsection{For the $5$-dimensional case}\label{sec32}
\begin{figure} [H]
	\centering
	\subfigure[]{
		\includegraphics[width=0.28\textwidth]{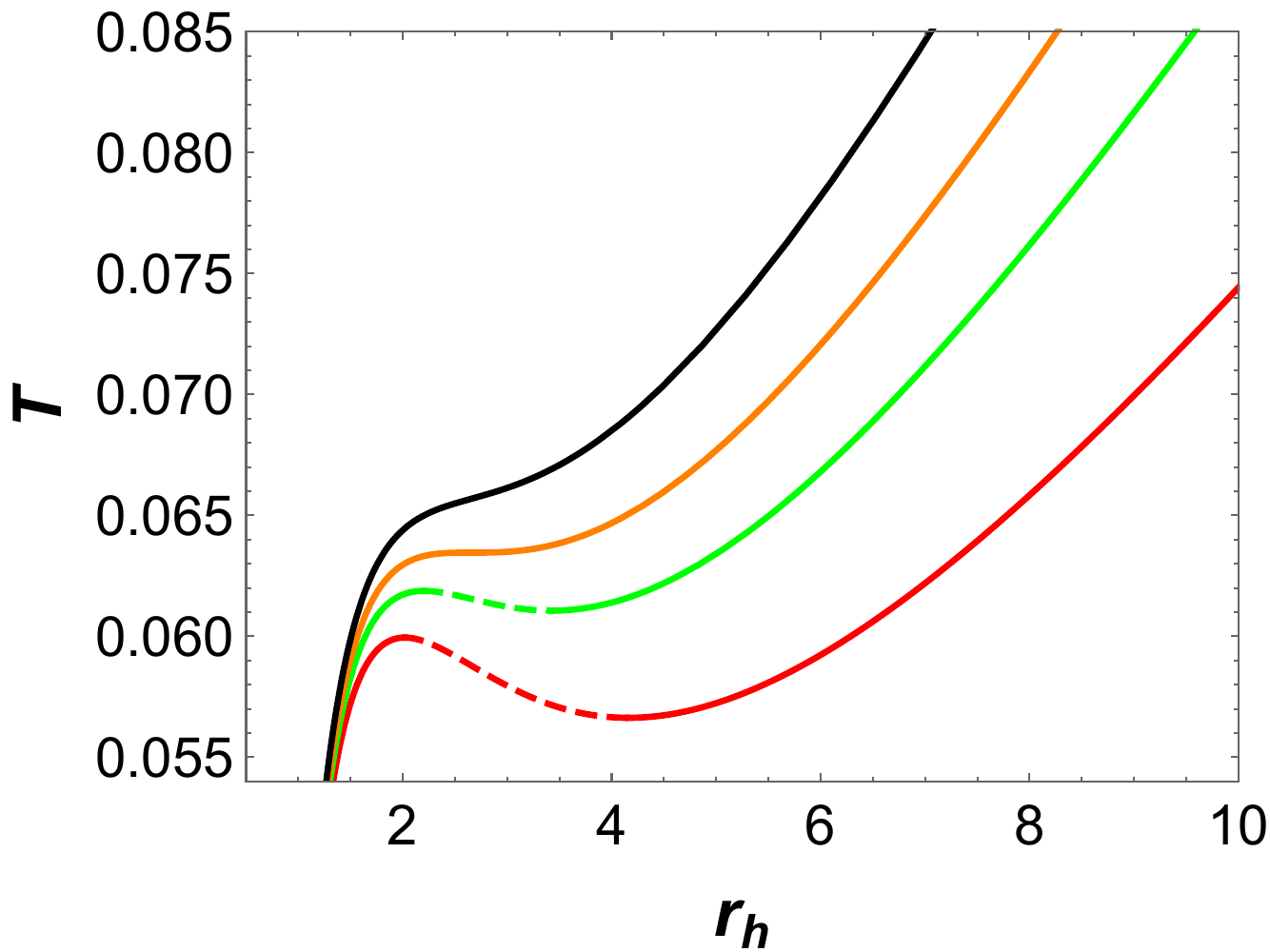}}
	\subfigure[]{
		\includegraphics[width=0.28\textwidth]{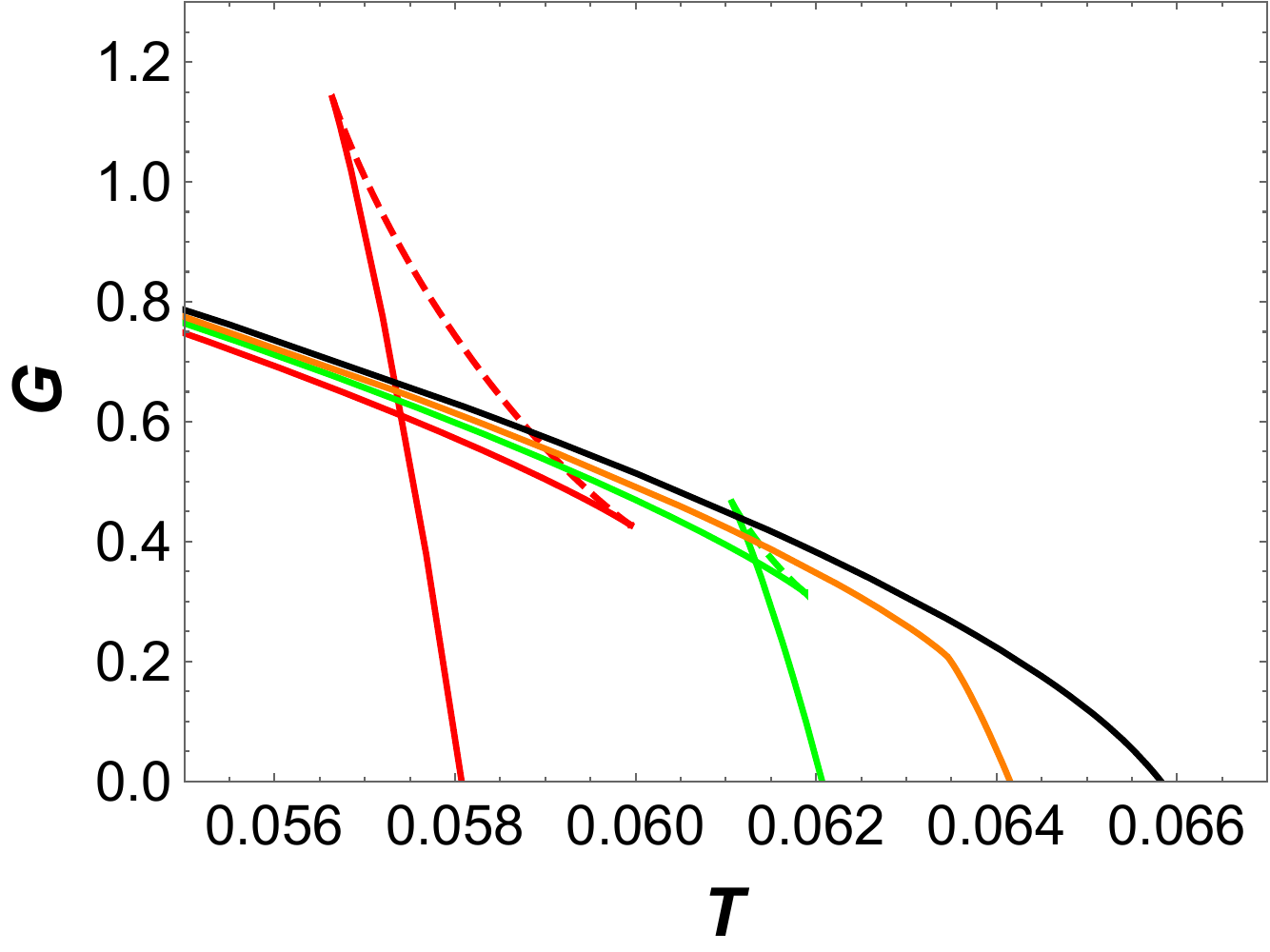}}
	\subfigure[]{
		\includegraphics[width=0.28\textwidth]{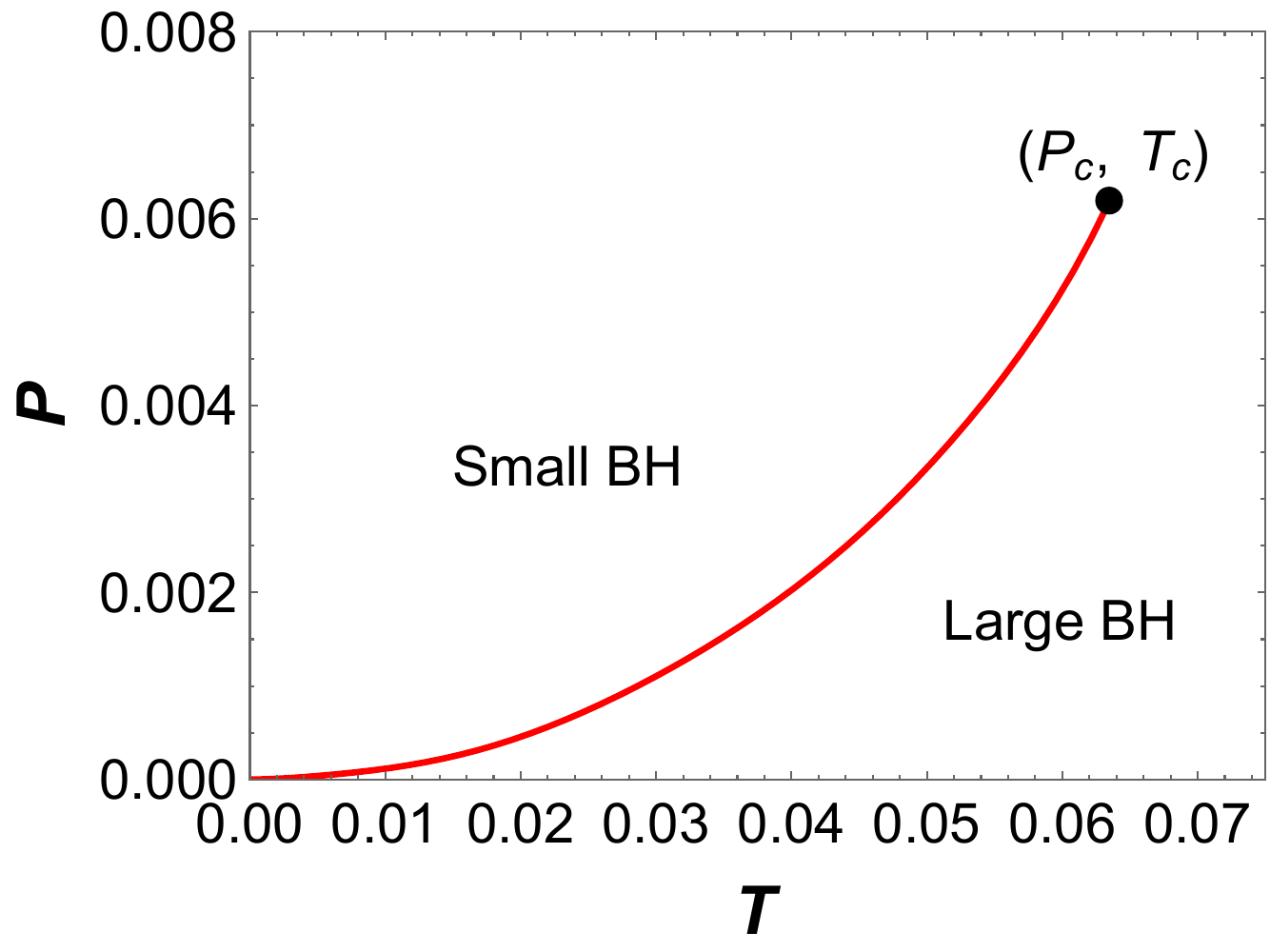}}
	\caption{\small (a) Temperature $T$ vs. $r_{h}$. (b) Gibbs free energy $G$ vs. $T$. The red, green, orange and black isobaric curves in (a) and (b) correspond to $P=0.0045$, $0.0055$, $0.00619309$ and $0.007$, respectively. (c) Phase diagram for the 5-dimensional dyonic AdS BH when $\alpha=0.5$ and $\beta=0.1$.}\label{fig2}
\end{figure}

In this subsection, we would like to study the phase transitions and phase diagrams of the 5-dimensional dyonic AdS BH. Similarly, the critical points for different combinations of parameters as shown in Tab.\ref{ta2}.
\begin{table}[H]
	\centering
	\captionsetup{font=footnotesize}
	{\caption {Critical points for different values of coupling constants $\alpha$ and $\beta$, when the dimension $D=5$.}
		{\footnotesize	\vspace{1mm}
			
			\begin{tabular} {ccc|ccc}
				\hline 
				{$\alpha$}  &{$\beta$}     &{$(T_{c},P_{c})$}  &{$\alpha$}  &{$\beta$}     &{$(T_{c},P_{c})$}   \\ 
				\hline
				{0.01}   &{0.01}   &{(0.1617879212, 0.03277316929)}    &{0.1}  &{0.01}   &{(0.1167835294, 0.01899568735)}      \\
				{0.01}   &{0.1}    &{(0.1617879212, 0.03277316929)}    &{0.1}  &{0.1}    &{(0.1167835294, 0.01899568735)}      \\
				{0.01}   &{0.5}    &{(0.1617879212, 0.03277316929)}    &{0.1}  &{0.5}    &{(0.1167835294, 0.01899568735)}       \\
				{0.01}   &{1}      &{(0.1617879212, 0.03277316929)}    &{0.1}  &{1}      &{(0.1167835294, 0.01899568735)}       \\
				{0.01}   &{10}     &{(0.1617879212, 0.03277316929)}    &{0.1}  &{10}     &{(0.1167835294, 0.01899568735)}      \\
				{0.01}   &{50}     &{(0.1617879212, 0.03277316929)}    &{0.1}  &{50}     &{(0.1167835294, 0.01899568735)}      \\
				\hline
				{0.5}	&{0.01}    &{(0.06346110823, 0.006193085877)}  &{1}    &{0.01}   &{(0.04564111900, 0.003251240167)}   \\
				{0.5}	&{0.1}     &{(0.06346110823, 0.006193085877)}  &{1}    &{0.05}   &{(0.04564111900, 0.003251240167)}   \\
				{0.5}	&{0.5}     &{(0.06346110823, 0.006193085877)}  &{1}    &{0.1}    &{(0.04564111900, 0.003251240167)}   \\
				{0.5}	&{1}       &{(0.06346110823, 0.006193085877)}  &{1}    &{0.5}    &{(0.04564111900, 0.003251240167)}   \\
				{0.5}	&{10}      &{(0.06346110823, 0.006193085877)}  &{1}    &{0.9}    &{(0.04564111900, 0.003251240167)}   \\
				{0.5}   &{50}      &{(0.06346110823, 0.006193085877)}  &{1}    &{1}      &{/}                                  \\
				\hline
				{10}	&{0.01}    &{(0.01452778369, 0.0003315037716)} &{50}   &{0.01}   &{(0.006497455295, 0.00006631401)}   \\
				{10}	&{0.05}    &{(0.01452778369, 0.0003315037716)} &{50}   &{0.05}   &{(0.006497455295, 0.00006631401)}   \\
				{10}	&{0.1}     &{(0.01452778369, 0.0003315037716)} &{50}   &{0.1}    &{(0.006497455295, 0.00006631401)}   \\
				{10}	&{0.5}     &{(0.01452778369, 0.0003315037716)} &{50}   &{0.5}    &{(0.006497455295, 0.00006631401)}   \\
				{10}	&{0.9}     &{(0.01452778369, 0.0003315037716)} &{50}   &{0.9}    &{(0.006497455295, 0.00006631401)}   \\
				{10}    &{1}       &{/}                                &{50}   &{1}      &{/}                                 \\
				\hline 
			\end{tabular}\label{ta2}}		
}\end{table}
As can be see from Tab.\ref{ta2}, there is only one critical point or no critical point for each choice of $\alpha$ and $\beta$.
Specifically, the critical point do not exist in the parameter region where both $\alpha$ and $\beta$ are all larger than or equal to $1$ ($\geqslant1$) simultaneously. And, when $\alpha$ is fixed, the critical point remains constant for different values of $\beta$.
But, it will decrease with the increase of the $\alpha$ for a fixed $\beta$.
For convenience, we chose $\alpha=0.5$ and $\beta=0.1$ as an example to study the phase transition of BHs. The critical point is given as
\begin{align}
	T_{c}=0.0634611,P_{c}=0.00619309.
\end{align}
The behaviors of temperature $T$ with respect to $r_{h}$ and Gibbs free energy $G$ with respect to $T$ are plotted in Fig.\ref{fig2}, respectively.

When $P< P_{c}$, there are two extremal points on both the red and green isobaric curves in the $T-r_{h}$ diagram, as well as two swallowtails along the isobaric curves in the $G-T$ plot, as shown in Figs.\ref{fig2} (a) and (b). As $P>P_{c}$, this leads to the disappearance of both extremal point of the temperature in the $T-r_{h}$ diagram and the swallowtail behavior in the $G-T$ plot, for their respective isobaric curves. The coexistence curve of the small and large BH phases is shown in Fig.\ref{fig2} (c). These results indicate a typical small/large BH phase transition, which is similar to the vdW liquid/gas phase transition. 
In addition, we have also found that the similar phase transitions always exist for other values of $\alpha$ and $\beta$ in Tab.\ref{ta2}. 

\subsection{For the $6$-dimensional case}\label{sec33}
In this subsection, we would like to study the phase transitions and phase diagrams of the 6-dimensional dyonic AdS BHs. Similarly, based on Eqs.(\ref{pressure}) and (\ref{conditions}), the critical points can be obtained which are shown in Tab.\ref{ta3}.
\begin{table}[H]
	\centering
	\captionsetup{font=footnotesize}
	{\caption {Critical points for different values of coupling constants $\alpha$ and $\beta$, when the dimension $D=6$.}
		{\footnotesize	\vspace{1mm}
			
			\begin{tabular} {ccccc}
				\hline 
				{$\alpha$}  &{$\beta$}     &{$(T_{c1},P_{c1})$}    &{$(T_{c2},P_{c2})$}     &{$(T_{c3},P_{c3})$}  \\ 
				\hline
				{0.1}   &{0.01}  &{(0.1389283596, 0.02790310620)}      &{/}     &{/}                  \\
				{0.1}   &{0.1}   &{(0.1389283596, 0.02790310620)}      &{/}     &{/}                 \\
				{0.1}   &{0.5}   &{(0.1389283596, 0.02790310620)}      &{/}     &{/}                 \\
				{0.1}   &{1}     &{(0.1389283596, 0.02790310620)}      &{/}     &{/}                 \\
				{0.1}   &{10}    &{(0.1389283596, 0.02790310620)}      &{/}     &{/}                 \\
				{0.1}   &{50}    &{(0.1389283596, 0.02790310620)}      &{/}     &{/}                 \\
				\hline
				{0.5}   &{0.01}  &{(0.06464171275, 0.006352744392)}    &{(0.06473618157, 0.006536184200)}   &{(0.06487209227, 0.00657574901)}  \\
				{0.5}   &{0.1}   &{(0.06464171275, 0.006352744392)}    &{(0.06473618157, 0.006536184200)}   &{(0.06487209227, 0.00657574901)}   \\
				{0.5}   &{0.5}   &{(0.06464171275, 0.006352744392)}    &{(0.06473618157, 0.006536184200)}   &{(0.06487209227, 0.00657574901)}   \\
				{0.5}   &{1}     &{(0.06464171275, 0.006352744392)}    &{(0.06473618157, 0.006536184200)}   &{(0.06487209227, 0.00657574901)}   \\
				{0.5}   &{10}    &{(0.06464171275, 0.006352744392)}    &{(0.06473618157, 0.006536184200)}   &{(0.06487209227, 0.00657574901)}   \\
				{0.5}   &{50}    &{(0.06464171275, 0.006352744392)}    &{(0.06473618157, 0.006536184200)}   &{(0.06487209227, 0.00657574901)}   \\	
				\hline 
				{1}     &{0.01}  &{(0.04592800869, 0.003308303581)}    &{(0.04593268003, 0.003311035759)}   &{(0.05592461935, 0.05061099927)}    \\
				{1}     &{0.1}   &{(0.04592800869, 0.003308303581)}    &{(0.04593268003, 0.003311035759)}   &{(0.05592461935, 0.05061099927)}  \\
				{1}     &{0.5}   &{(0.04592800869, 0.003308303581)}    &{(0.04593268003, 0.003311035759)}   &{(0.05592461935, 0.05061099927)}     \\
				{1}     &{1}     &{/}                                  &{/}                                 &{/}                 \\
				\hline
				{10}    &{0.01}  &{(0.01452878785, 0.0003315722183)}   &{(0.01452878790, 0.0003315722265)}  &{(0.07964136252, 9.372787591)}   \\
				{10}    &{0.1}   &{(0.01452878785, 0.0003315722183)}   &{(0.01452878790, 0.0003315722265)}  &{(0.07964136252, 9.372787591)}    \\
				{10}    &{0.5}   &{(0.01452878785, 0.0003315722183)}   &{(0.01452878790, 0.0003315722265)}  &{(0.07964136252, 9.372787591)}     \\
				{10}    &{1}     &{/}      &{/}     &{/}                 \\
				\hline
				{50}    &{0.01}  &{(0.00649747332857, 0.0000663145587)} &{(0.00649747332858, 0.000066314558701)} &{(0.1182077162, 238.3354046)}    \\
				{50}    &{0.1}   &{(0.00649747332857, 0.0000663145587)} &{(0.00649747332858, 0.000066314558701)} &{(0.1182077162, 238.3354046)}     \\
				{50}    &{0.5}   &{(0.00649747332857, 0.0000663145587)} &{(0.00649747332858, 0.000066314558701)} &{(0.1182077162, 238.3354046)}    \\
				{50}    &{1}     &{/}      &{/}     &{/}                 \\
				\hline
			\end{tabular}\label{ta3}}		
}\end{table}
Comparing Tab.\ref{ta3} with Tabs.\ref{ta1} and \ref{ta2}, it can be see that the results in dimension $D=6$ are more interesting than those in dimensions $D=4$ and $5$. 
As $\alpha = 0.1$, the system only exists one critical point whatever $\beta$ is. But, three critical points can be observed for $\alpha\geqslant0.5$ in Tab.\ref{ta3}. In the parameter region where $\alpha\geqslant1$ and $\beta\geqslant1$ simultaneously, all critical points disappear.
It is easy to conclude that $\alpha$ has some stronger effect on the critical points by comparing with the $\beta$.
Therefore, in general, we take $\alpha=0.1$ and $\beta=0.1$, $\alpha=0.5$ and $\beta=0.1$ and $\alpha=1$ and $\beta=0.1$ as three typical examples to study the phase transitions and phase diagrams of BHs in dimension $D=6$, and further discuss the influence of $\alpha$ and $\beta$ on BH phase transitions.

\subsubsection{Example I: $\alpha=0.1$ and $\beta=0.1$}\label{sec331}
\begin{figure} [H]
	\centering
	\subfigure[]{
		\includegraphics[width=0.28\textwidth]{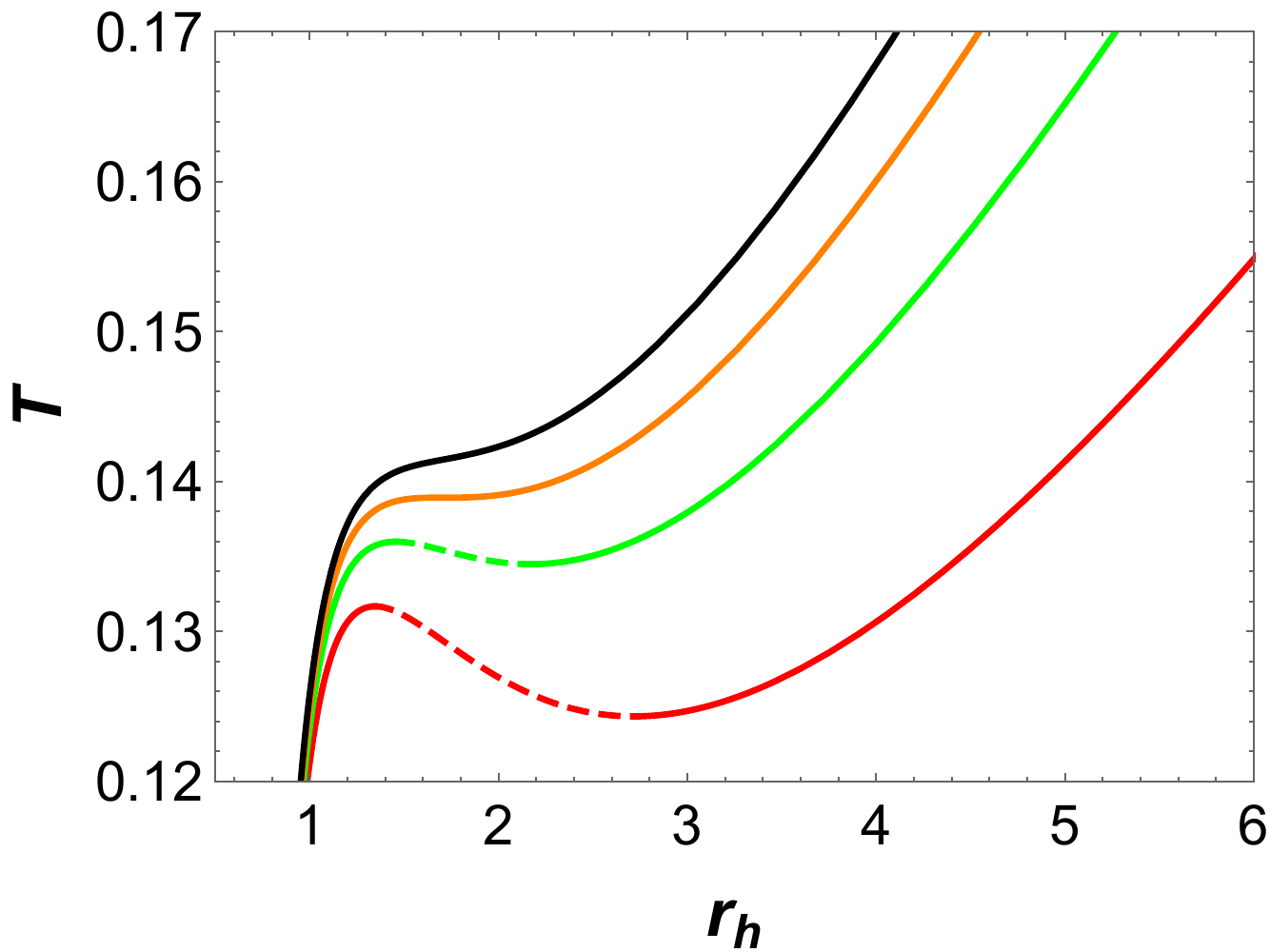}}
	\subfigure[]{
		\includegraphics[width=0.28\textwidth]{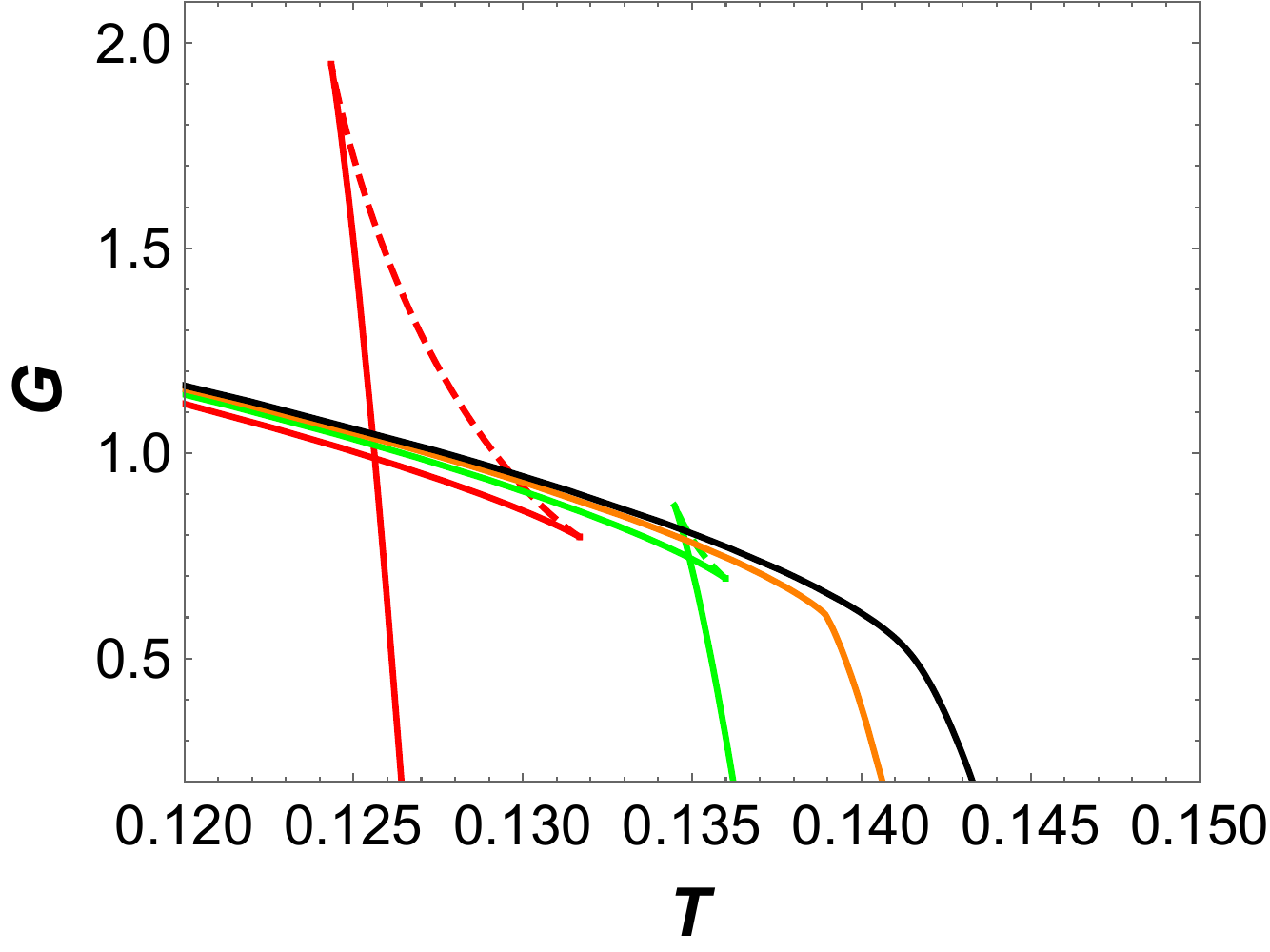}}
	\subfigure[]{
		\includegraphics[width=0.28\textwidth]{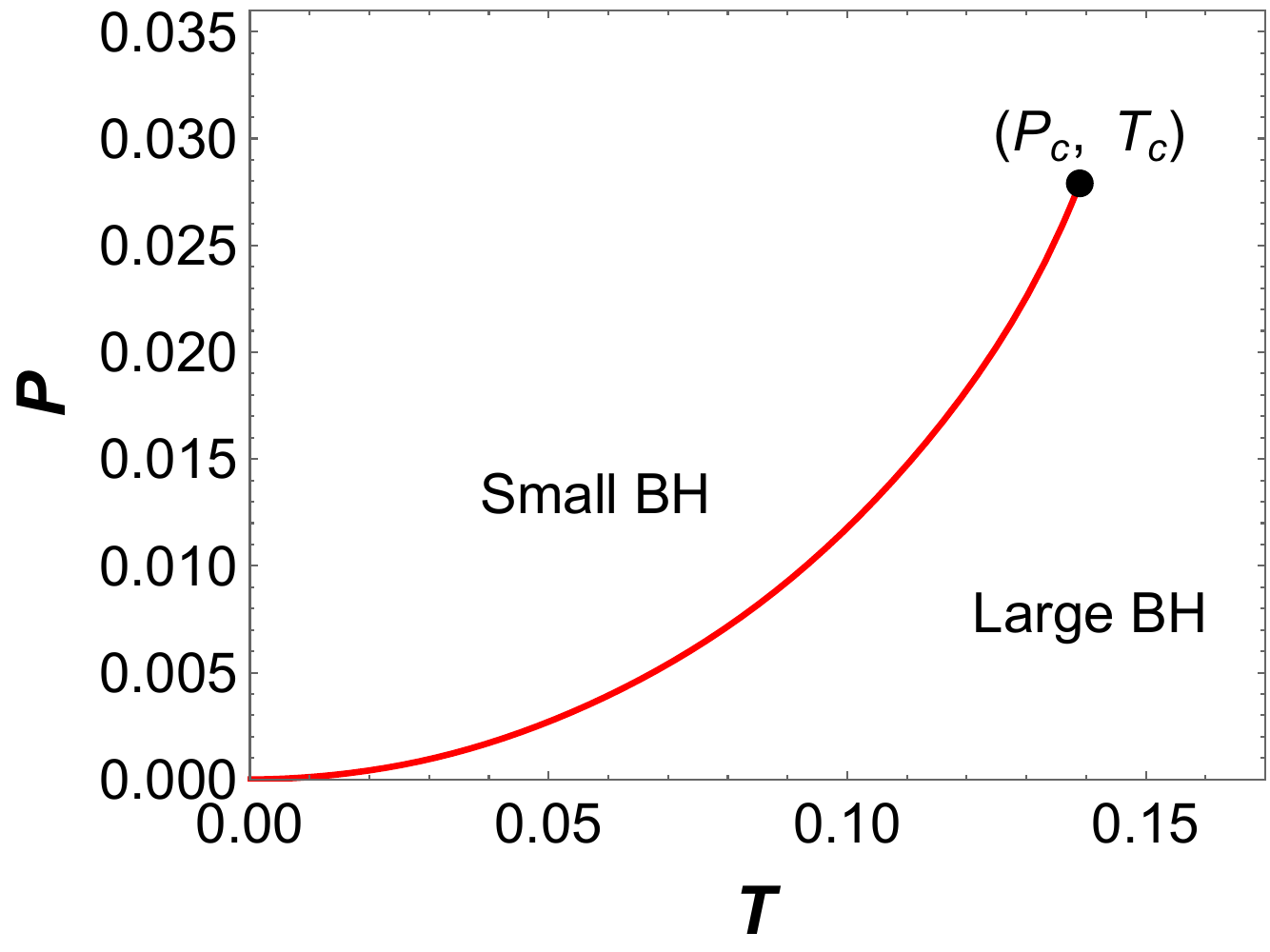}}
	\caption{\small (a) Temperature $T$ vs. $r_{h}$. (b) Gibbs free energy $G$ vs. $T$. The red, green, orange and black isobaric curves in (a) and (b) correspond to $P=0.02$, $0.025$, $0.0279031$ and $0.03$, respectively. (c) Phase diagram for the 6-dimensional dyonic AdS BHs when $\alpha=0.1$ and $\beta=0.1$.}\label{fig3}
\end{figure}

In this subsubsection, when $\alpha=0.1$ and $\beta=0.1$, the critical point can be obtained, it is 
\begin{align}
	T_{c}=0.138928,P_{c}=0.0279031.
\end{align}
Figs.\ref{fig3} (a) and (b) we plotted are devoted to illustrate the behavior of temperature $T$ vs. $r_{h}$ and Gibbs free energy $G$ vs. $T$, respectively. From these figures, it can bee see that when $\alpha=0.1$ and $\beta=0.1$ the BH system only undergoes a small/large BH phase transition. Also, we constructed a phase diagram shown in Fig.\ref{fig3}(c).
It is easy to find that the phase transitions and phase diagrams in this case is similar to that obtained in dimensions $D=4, 5$.

\begin{figure} [H]
	\centering
	\subfigure[]{
		\includegraphics[width=0.34\textwidth]{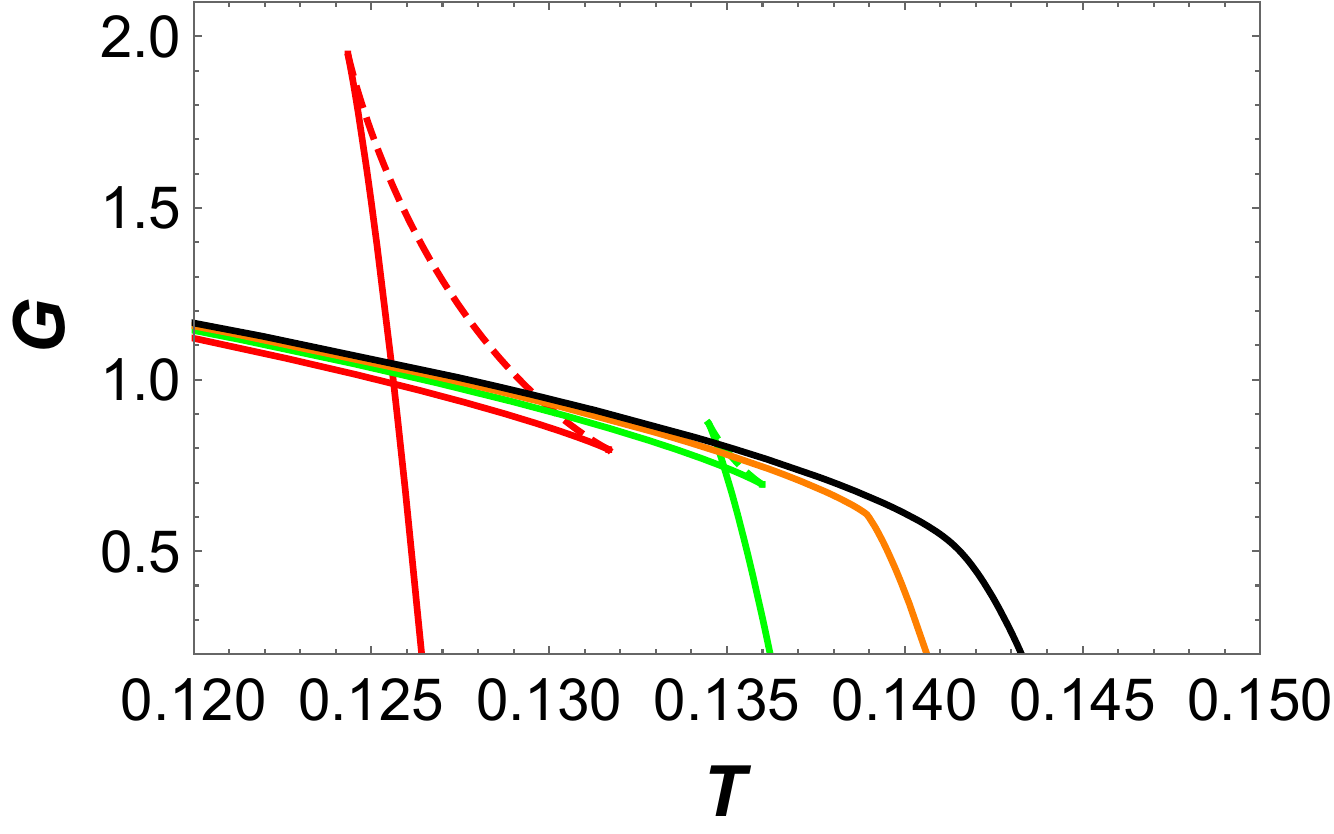}}
	\subfigure[]{
		\includegraphics[width=0.34\textwidth]{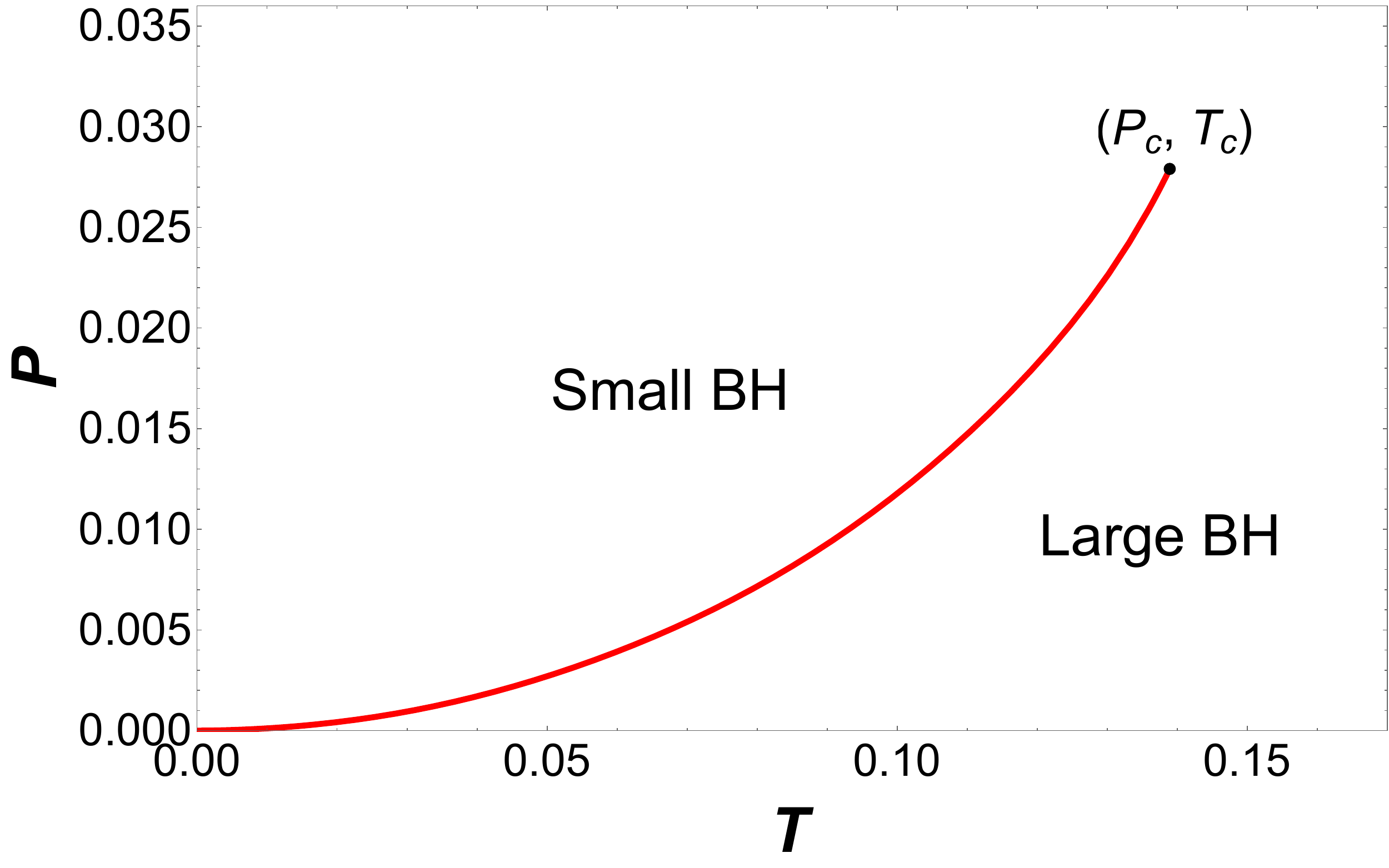}}
	\caption{\small (a) $G$ vs. $T$. (b) Phase diagram for the 6-dimensional dyonic AdS BHs when $\alpha=0.1$ and $\beta=0.5$.}\label{fig4}
\end{figure}

Furthermore, we also plotted the $G-T$ and $P-T$ diagrams for coupling constants $\alpha=0.1$ and $\beta=0.5$ as shown in Fig.\ref{fig4}. It can be see that the system undergoes a small/large BH phase transition in this parameter region. By comparing Fig.\ref{fig4} with Fig.\ref{fig3}, it can be observed that they are almost identical. This implies that the parameter $\beta$ has a little effect on the BH phase transition in this case.

\subsubsection{Example II: $\alpha=0.5$ and $\beta=0.1$}\label{sec332}
\begin{figure} [H]
	\centering
	\subfigure[]{
		\includegraphics[width=0.4\textwidth]{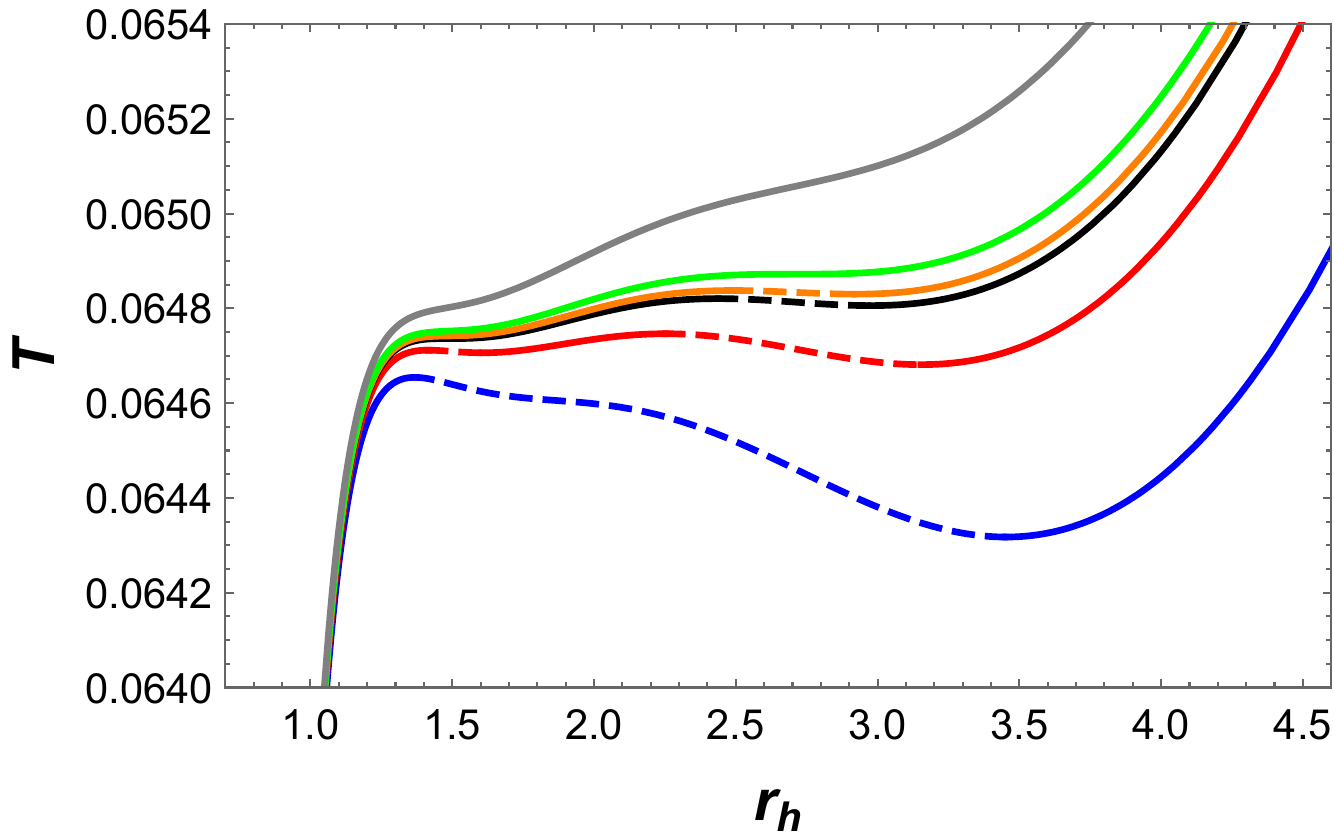}}
	\subfigure[]{
		\includegraphics[width=0.4\textwidth]{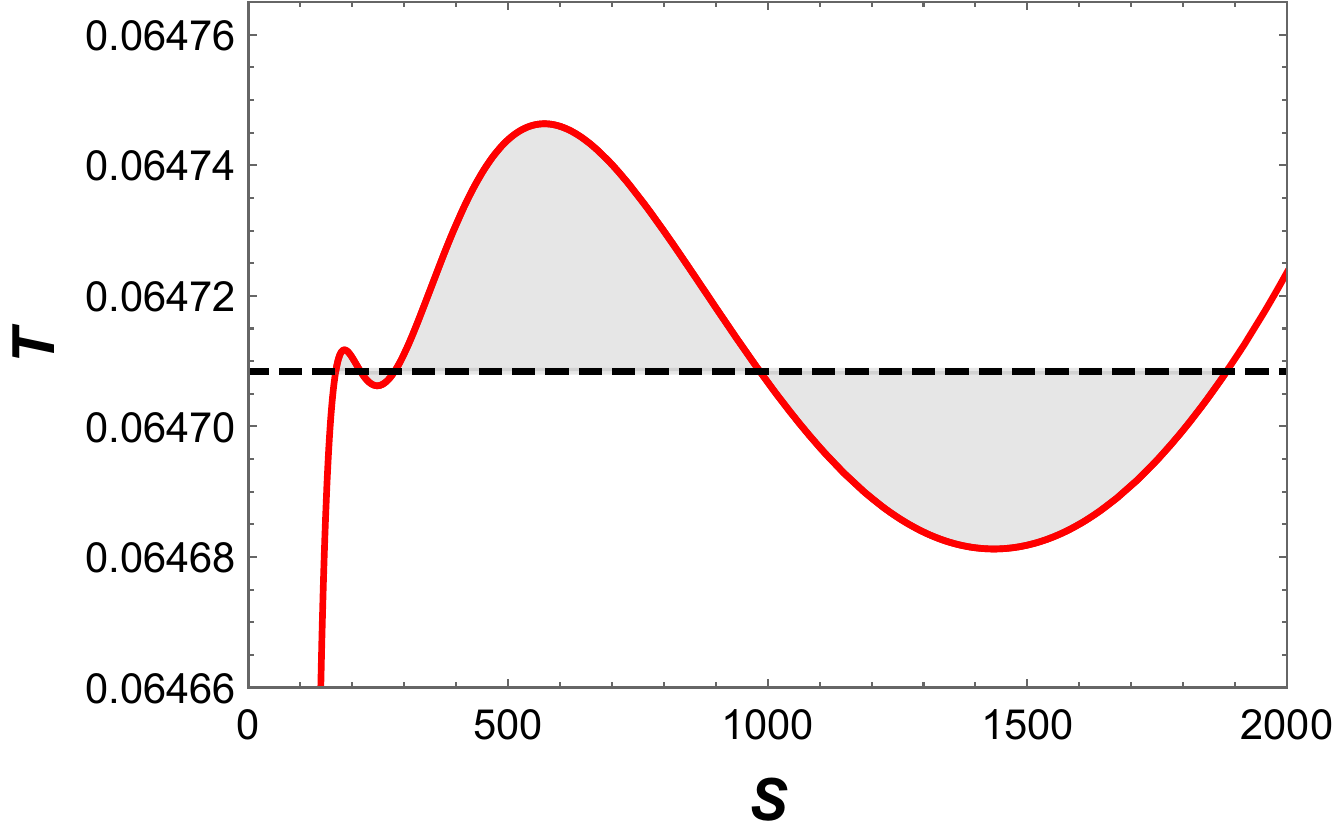}}
	\caption{\small(a) $T$ vs. $r_{h}$. The blue, red, black, orange, green and gray isobaric curves correspond to $P=0.0063$, $0.00646982$, $0.00653618$, $0.00655$, $0.00657575$ and $0.0067$, respectively. (b) $T-S$ diagram of two pairs of equal area regions at pressure $P = P_{t} = 0.00646982$. The horizontal line has a temperature $T = T_{t} = 0.0647085$.
}\label{fig5}
\end{figure}

In this subsubsection, we set $\alpha=0.5$ and $\beta=0.1$ to study the phase transitions and phase diagrams of the BHs. 
In this case, three critical points can be obtained, they are
\begin{align}
	T_{c1}=0.0646417,P_{c1}=0.00635274,\\
	T_{c2}=0.0647362,P_{c2}=0.00653618,\\
	T_{c3}=0.0648721,P_{c3}=0.00657575.
\end{align}
Firstly, we will analyze the behavior of temperature $T$ with respect to $r_{h}$. In Fig.\ref{fig5}(a), the temperature $T$ is plotted as a function of $r_{h}$, where pressure is considered as a variable. 
When $P<P_{c1}$, the blue isobaric curve on the $T-r_{h}$ diagram has two extremal points that divide it into three branches: the stable small BH branch, the unstable intermediate BH branch, and the stable large BH branch.
For $P_{c1}<P<P_{c2}$, the red isobaric curve has four extremal points that divide it into five branches: the stable small BH branch, the unstable small BH branch, the stable intermediate BH branch, the unstable large BH branch, and the stable large BH branch.
 For the red isobaric curve, one can utilize Maxwell equal area laws to plot two pairs of equal area regions at pressure $P=P_{t}=0.00646982$, which is shown in Fig.\ref{fig5}(b). It should be noted that the two pairs of regions obtained by using Maxwell equal area laws have the same temperature $T=T_{t}=0.0647085$. This indicates that the BH system undergoes two phase transitions at the same pressure and temperature. This result implies the existence of the triple point, where small, intermediate, and large BH phases can coexist simultaneously. 
When the pressure reaches $P_{c2}$, the BH system undergoes a second-order intermediate/large BH phase transition.

For $P_{c2}<P<P_{c3}$, two extremal points appear on the orange isobaric curve. As a result, three distinct branches can be observed: the stable intermediate BH branch, the unstable intermediate BH branch, and the stable large BH branch. This result suggests the occurrence of the intermediate/large BH phase transition in this pressure range. However, this first-order phase transition turns to a second-order phase transition as the pressure approaches $P_{c3}$. For $P>P_{c3}$, there are no extremal point on the gray isobaric curve, indicating that temperature $T$ becomes a monotonic function of $r_{h}$, and thus there exists only one BH branch.

\begin{figure} [H]
	\centering
	\subfigure[]{
		\includegraphics[width=0.4\textwidth]{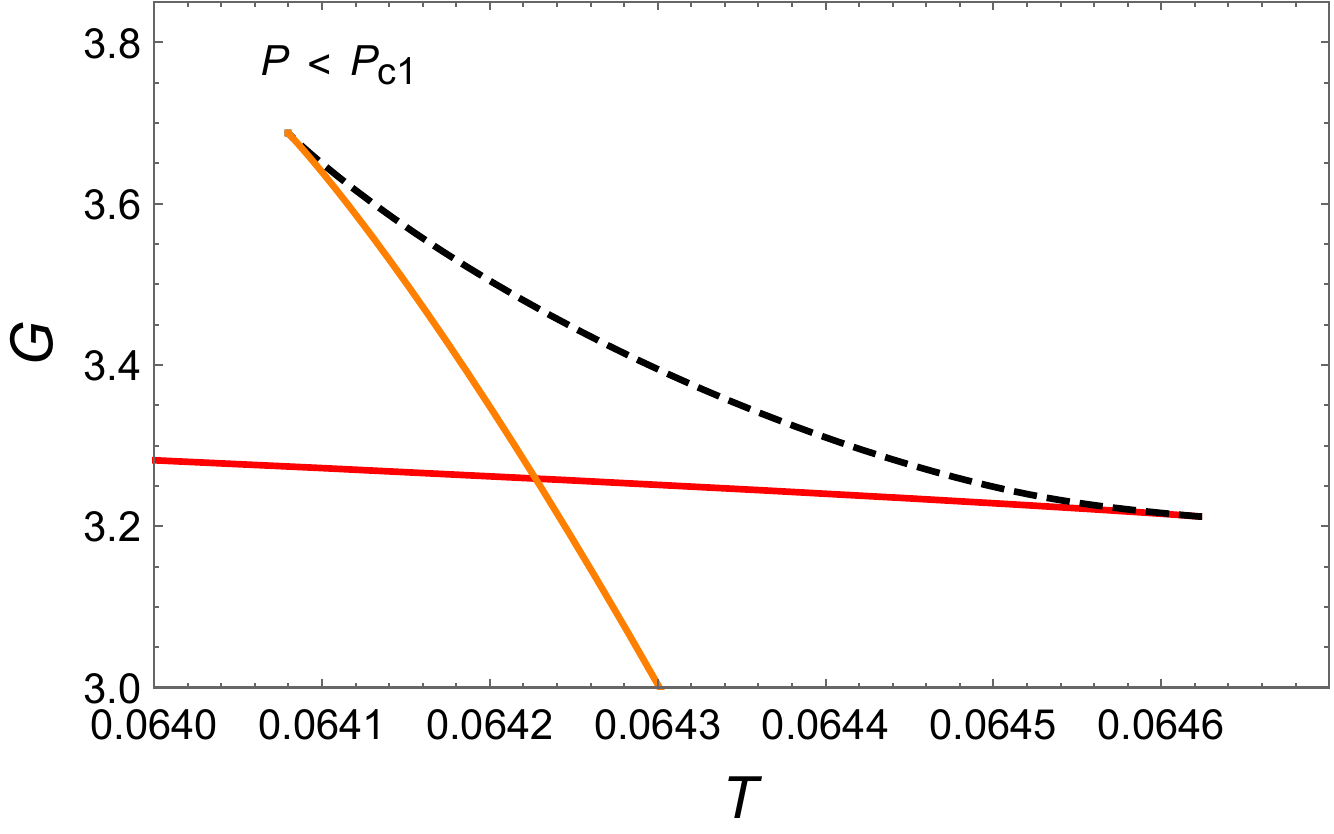}}
	\subfigure[]{
		\includegraphics[width=0.4\textwidth]{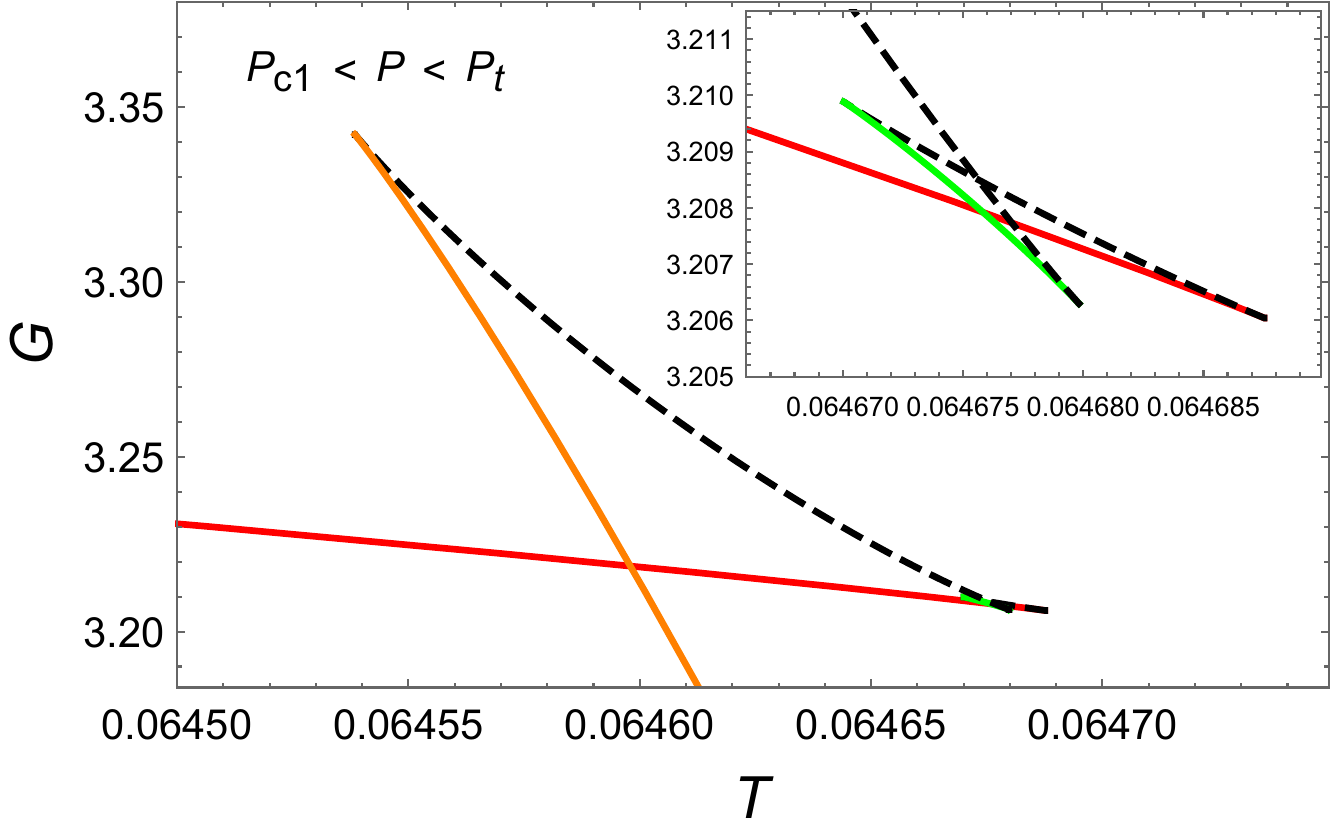}}
		\subfigure[]{
		\includegraphics[width=0.4\textwidth]{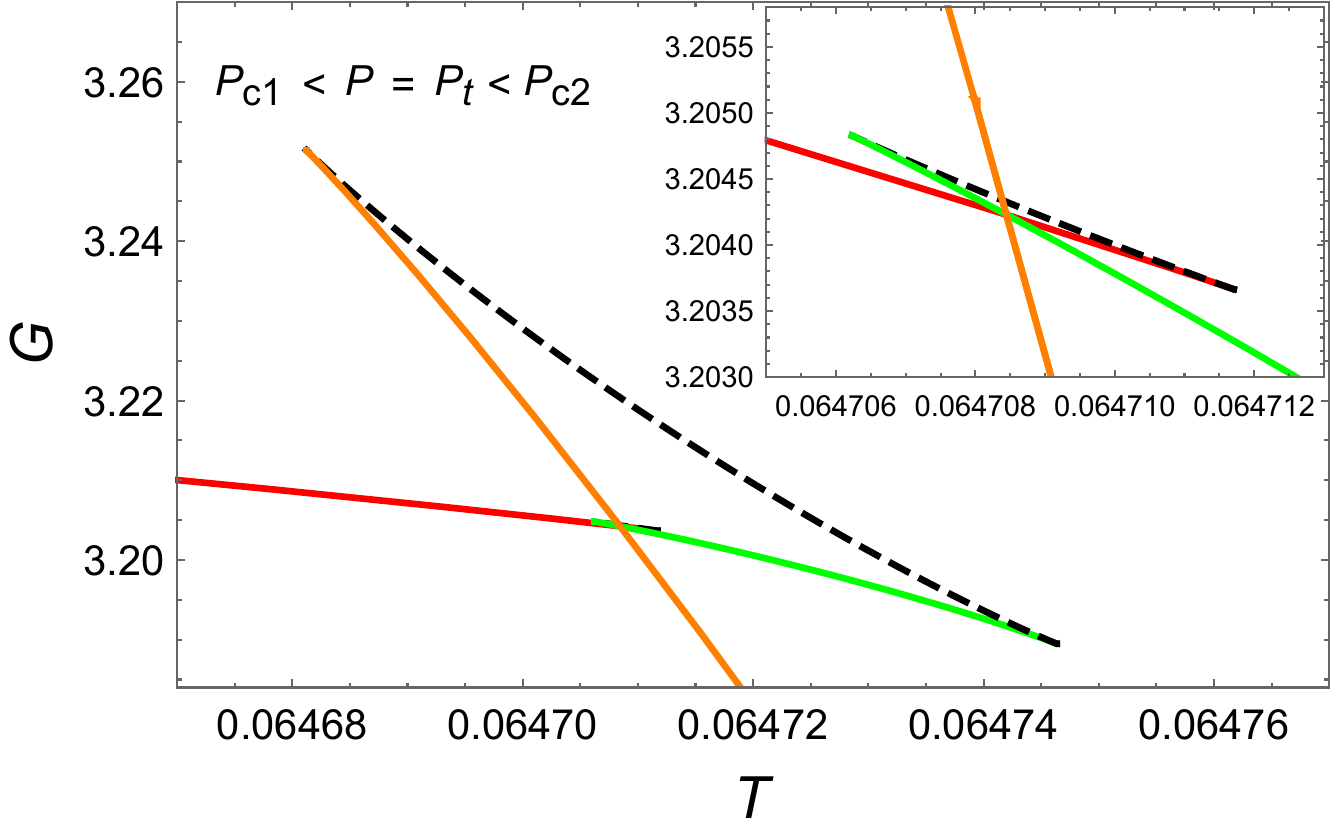}}
		\subfigure[]{
		\includegraphics[width=0.4\textwidth]{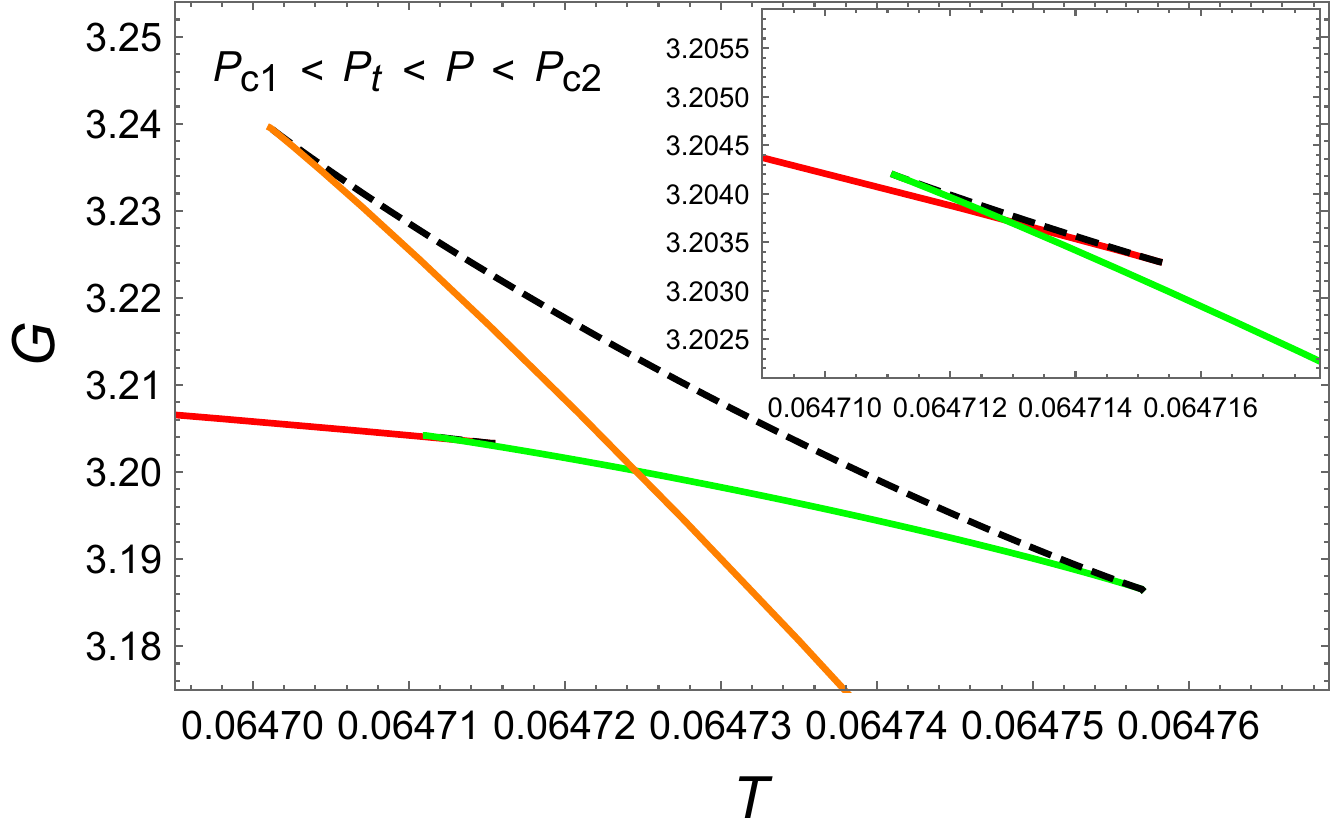}}
		\subfigure[]{
		\includegraphics[width=0.4\textwidth]{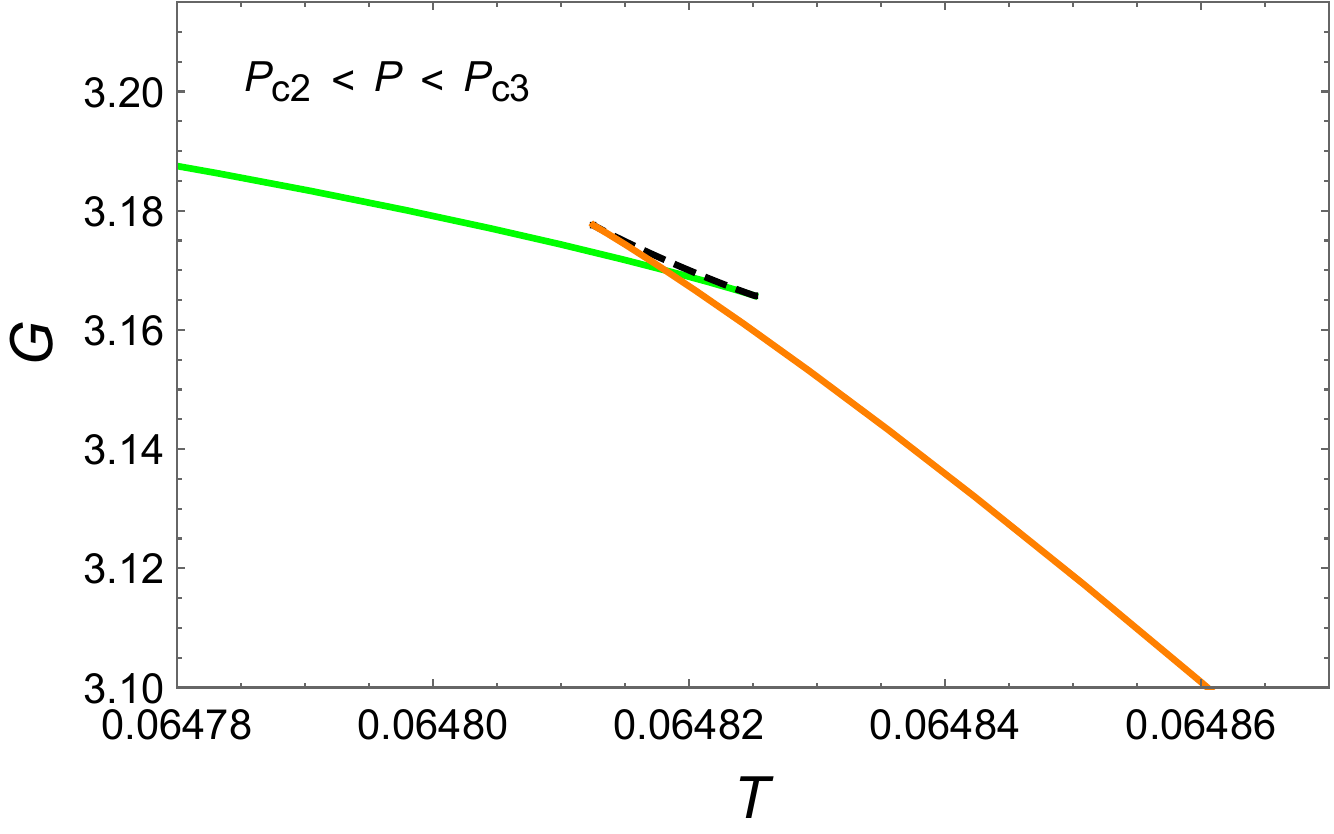}}
		\subfigure[]{
		\includegraphics[width=0.4\textwidth]{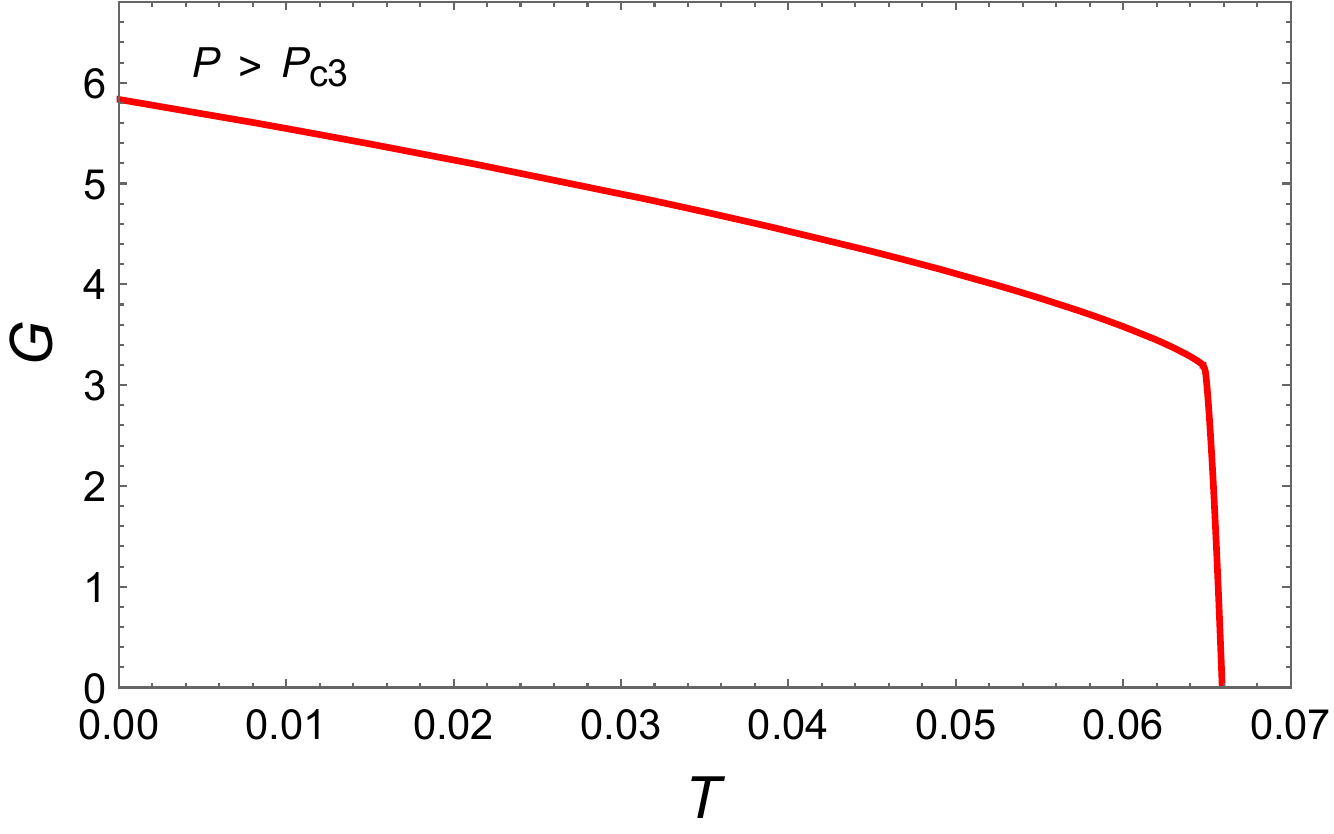}}
	\caption{\small 
		$G$ vs. $T$. The red, green and orange curves represent small, intermediate and large BHs, respectively. (a) $P=0.0062$. (b) $P=0.0064$. (c) $P=P_{t}=0.00646982$. (d) $P=0.00648$. (e) $P=0.00654$. (f) $P=0.0066$.}\label{fig6}
\end{figure}

Next, we will analyze the behavior of Gibbs free energy $G$ with respect to $T$. Here, we have plotted the behavior of Gibbs free energy in Fig.\ref{fig6}. It is important to note that the nonsmooth points on the isobaric curves in the $G-T$ diagram correspond to the extremal points on the isobaric curves in the $T-r_{h}$ diagram. For $P<P_{c1}$, there is a swallowtail behavior in Fig.\ref{fig6}(a), suggesting the existence of a small/large BH phase transition. As the pressure increased to the range of $P_{c1}<P<P_{t}$, two swallowtail behaviors emerged in Fig.\ref{fig6}(b), which indicates the existence of three stable BH branches, and there seems to be two phase transitions. 
However, one of them will be suppressed and does not participate in the phase transition since its higher free energy. 
Therefore, there is only a small/large BH phase transition as the pressure in the region $P_{c1}<P<P_{t}$. When $P=P_{t}$, as shown in Fig.\ref{fig6}(c), the intersection of the two swallowtails occurs at the same position($P_{t},T_{t}$). This indicates that the three stable BH branches intersect at this point. Therefore, this result suggests that the triple point $(T_{t}, P_{t})$ exists for the current situation in this system, which means the small, intermediate and large BH phases can coexist simultaneously. 
For the case $P_{t}<P<P_{c2}$, as shown in Fig.\ref{fig6}(d), we find that two swallowtails are also observed, and all three BH branches can participate in the phase transitions. This implies that at the same pressure but different temperatures, both the small/intermediate BH phase transition and the intermediate/large BH phase transition can occur simultaneously. 
As $P_{c2}<P<P_{c3}$, from Fig.\ref{fig6}(e), it shows that there is only one swallowtail, which indicates that only the intermediate/large BH phase transition presents. However, for $P>P_{c3}$, the Gibbs free energy $G$ becomes a monotonic function of $T$, thereby no phase transition occurs in the system.

\begin{figure} [H]
	\centering
	\subfigure[]{
		\includegraphics[width=0.4\textwidth]{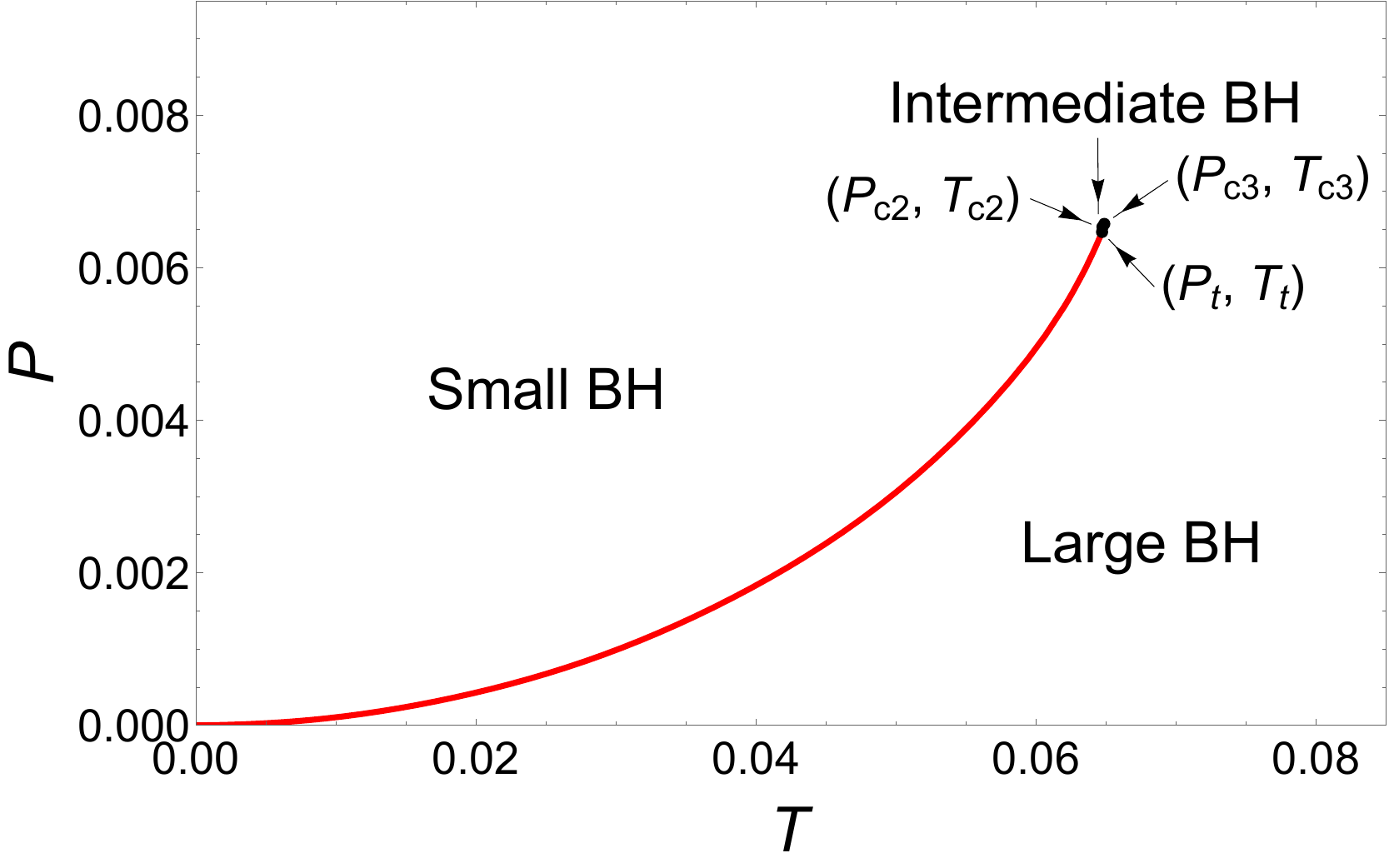}}
	\subfigure[]{
		\includegraphics[width=0.4\textwidth]{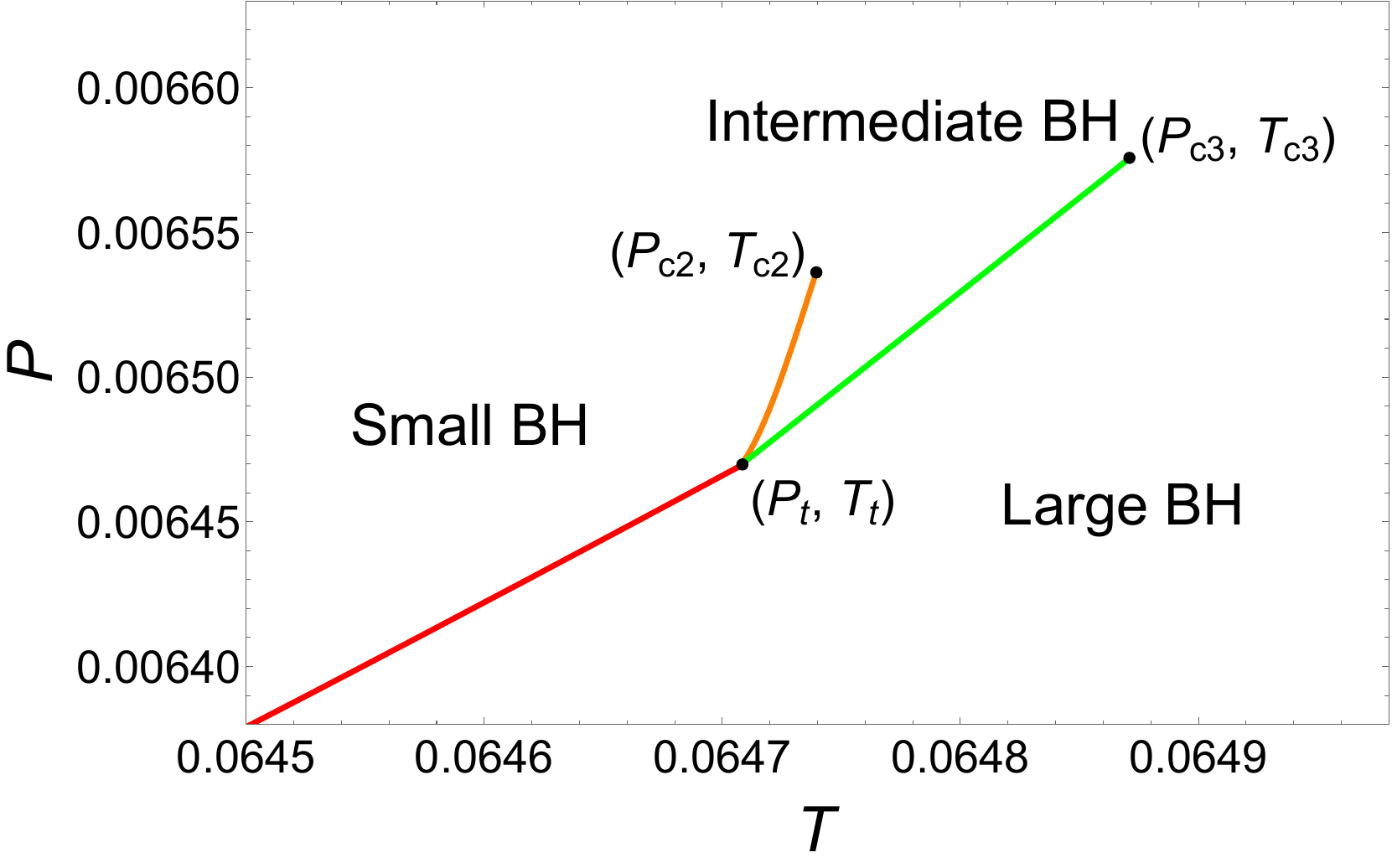}}
	\caption{\small Phase diagram for the 6-dimensional dyonic AdS BHs when $\alpha=0.5$ and $\beta=0.1$. (a) Entire phase diagram. (b) An enlarged view near the triple point ($P_{t}, T_{t}$).
	}\label{fig7}
\end{figure}

Finally, we constructed the $P-T$ phase diagram as shown in Fig.\ref{fig7}. Specifically, Fig.\ref{fig7}(a) is the global phase diagram, while Fig.\ref{fig7}(b) is a local one near the triple point ($P_{t}, T_{t}$).
From the $P-T$ phase diagram, it can be observed in current situation that when $P<P_{t}$, the system undergoes a small/large BH phase transition. At $P=P_{t}$, the small, intermediate and large BH phases can coexist simultaneously. In the range $P_{t}<P<P_{c2}$, the system undergoes both the small/intermediate BH phase transition and the intermediate/large BH phase transition simultaneously. For $P_{c2}<P<P_{c3}$, the system only undergoes the intermediate/large BH phase transition.

\begin{figure} [H]
	\centering
	\subfigure[]{
		\includegraphics[width=0.38\textwidth]{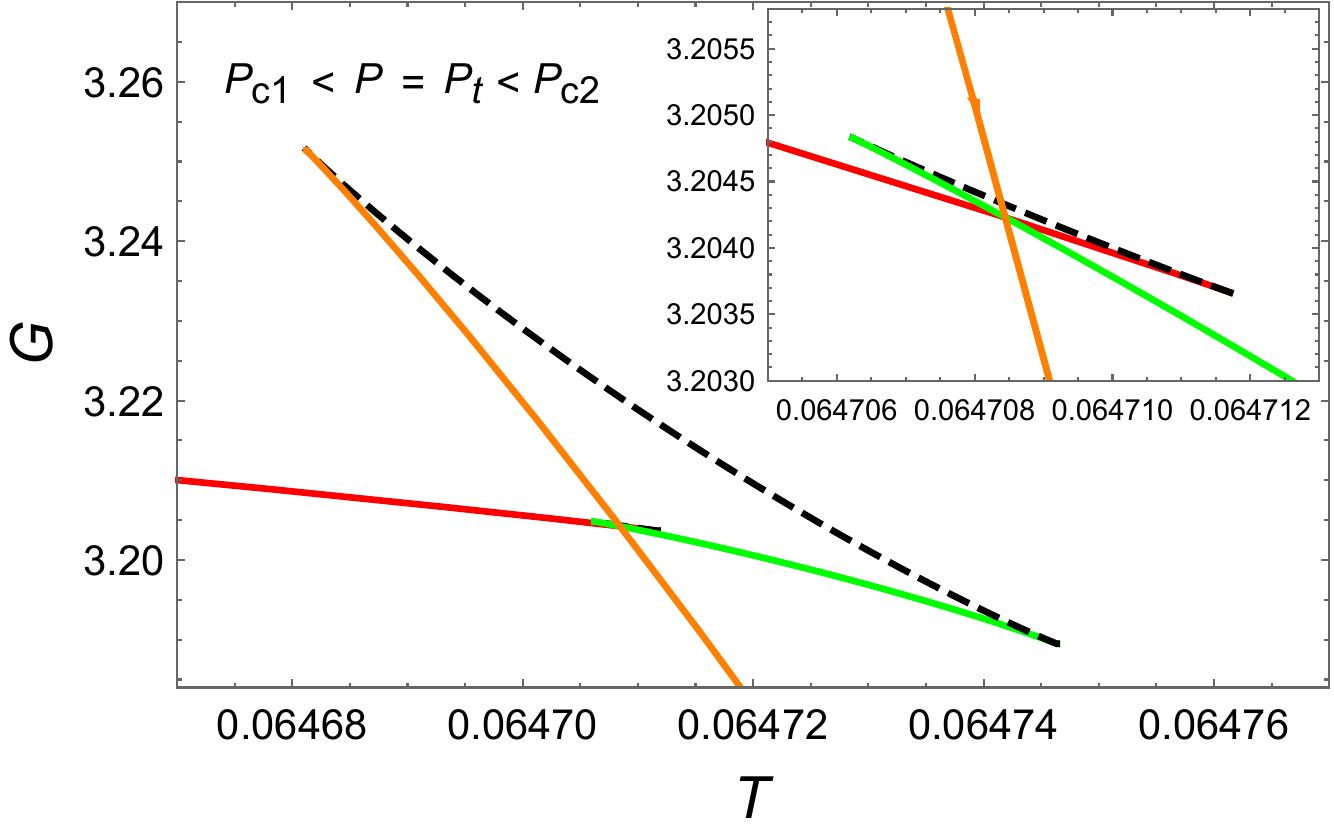}}
	\subfigure[]{
		\includegraphics[width=0.38\textwidth]{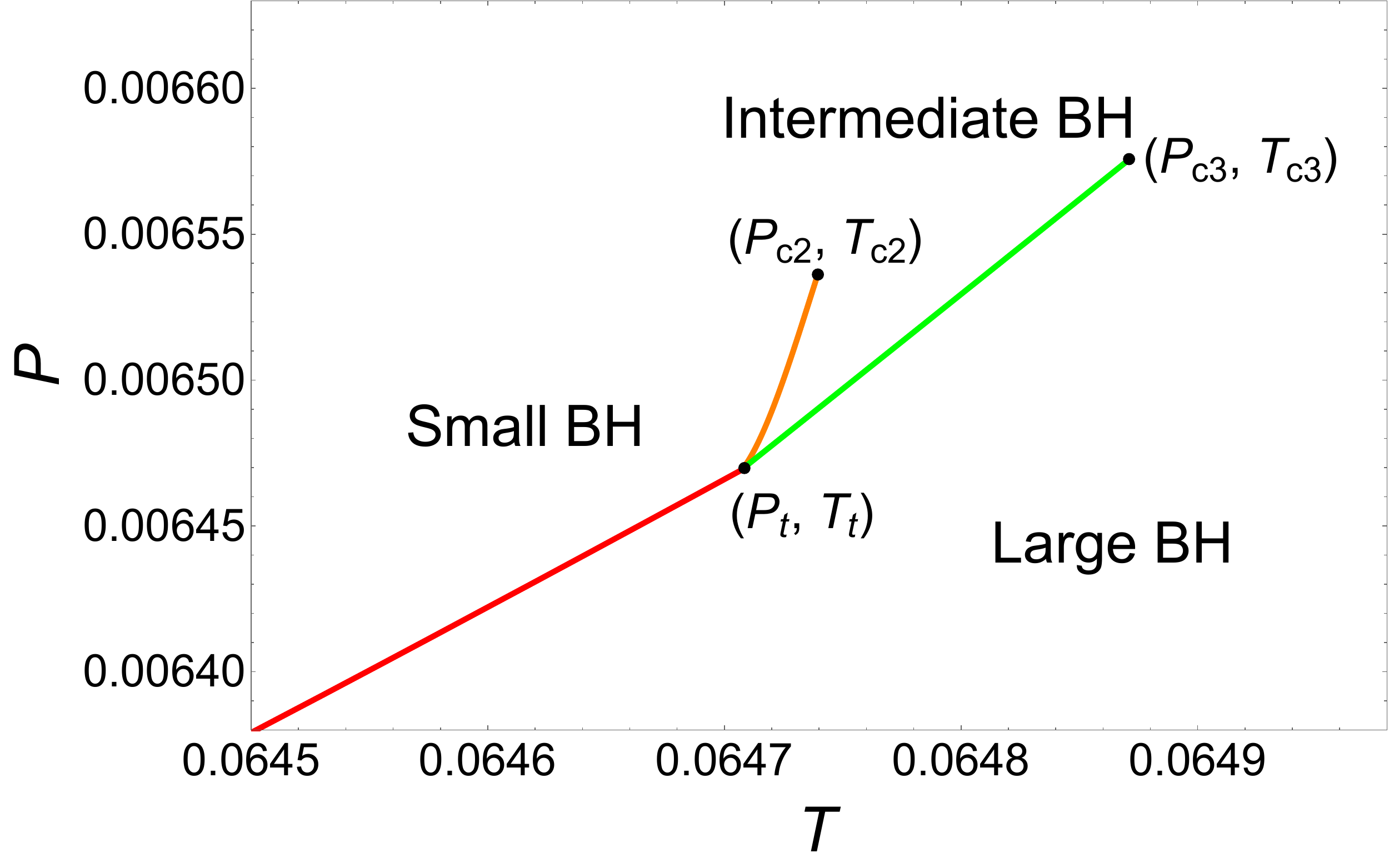}}
	\caption{\small (a) $G$ vs. $T$. (b) Phase diagram for the 6-dimensional dyonic AdS BH when $\alpha=0.5$ and $\beta=50$.
	}\label{fig8}
\end{figure}

In addition, we also investigated the phase transitions of BH for $\alpha=0.5$ and $\beta=50$. 
The results indicate that the triple point ($P_t=0.006469822191382499, T_t=0.06470845191069557$) for $\alpha=0.5$ and $\beta=50$ is almost identical to that obtained for $\alpha=0.5$ and $\beta=0.1$ ($P_t=0.006469822191382416, T_t=0.06470845191069545$), with an accuracy  up to $13$ significant digits. And, we have plotted $G-T$ diagram at the triple point and $P-T$ phase diagram near the triple point for $\alpha=0.5$ and $\beta=50$, as illustrated in Fig.\ref{fig8}. By comparing Figs.\ref{fig6}, \ref{fig7} with \ref{fig8}, there is almost no difference for the $G-T$ and $P-T$ plots that can be observed. This finding implies that when $\alpha=0.5$, the BH system always posses remarkably similar phase transitions, irrespective of the variations of $\beta$. This means the parameter $\alpha$ plays a key role during the studying of the phase transition of BH in dimension $D=6$.
 
\subsubsection{Example III: $\alpha=1$ and $\beta=0.1$}\label{sec333}
In this subsubsection, by taking $\alpha=1$ and $\beta=0.1$, there also are three critical points, which is
\begin{align}
	T_{c1}=0.0459280,P_{c1}=0.00330830,\\
	T_{c2}=0.0459327,P_{c2}=0.00331104,\\
	T_{c3}=0.0559246,P_{c3}=0.05061100.
\end{align}

\begin{figure} [H]
	\centering
	\subfigure{
		\includegraphics[width=0.4\textwidth]{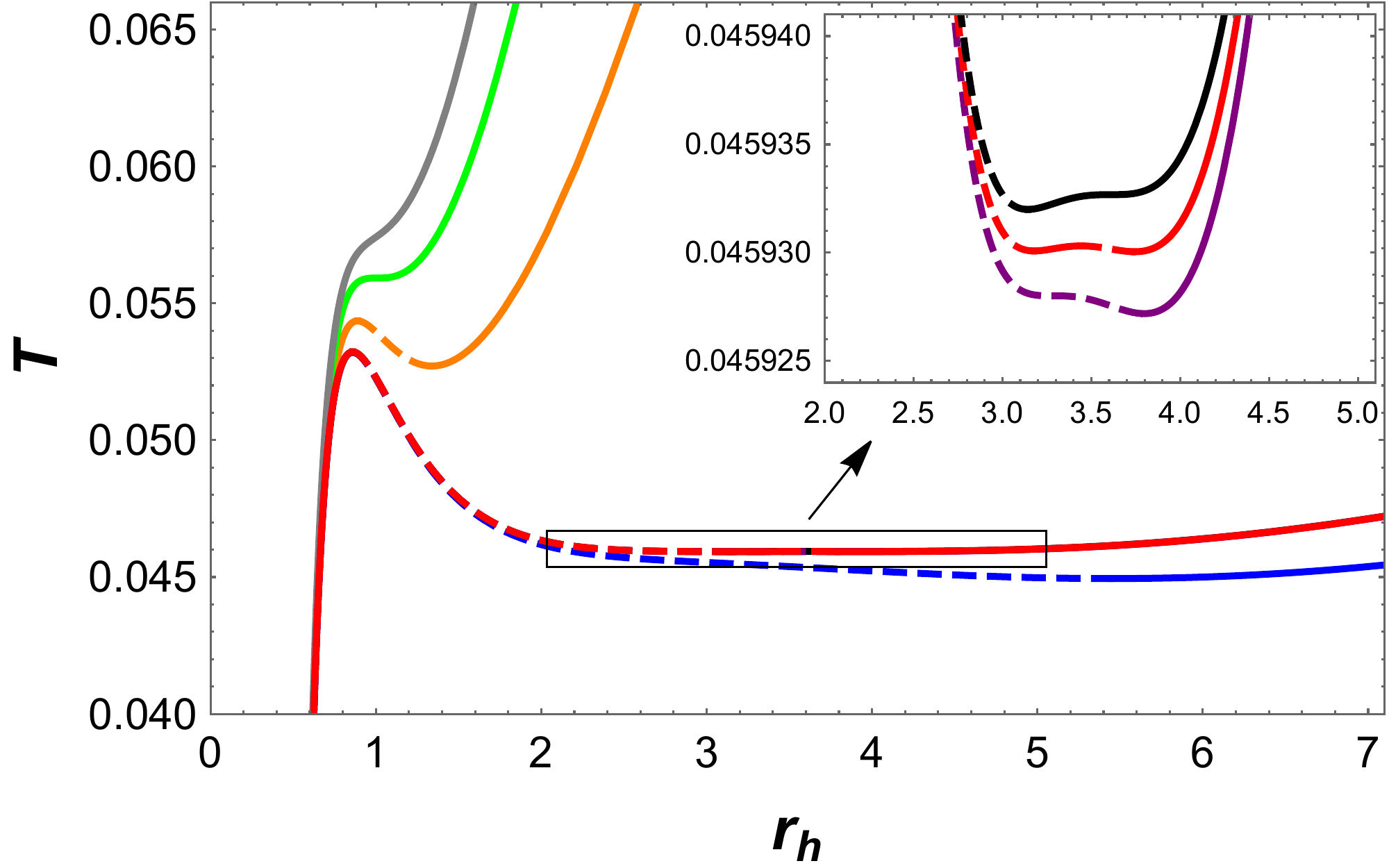}}
	\caption{\small Temperature $T$ vs. $r_{h}$ for $P=0.003$ (blue curve), $P=0.0033083$ (purple curve), $P=0.0033097$ (red curve), $P=0.00331104$ (black curve), $P=0.025$ (orange curve), $P=0.050611$ (green curve) and $P=0.07$ (gray curve).
		}\label{fig9}
\end{figure}

Firstly, we plotted the temperature $T$ with respect to $r_{h}$ in Fig.\ref{fig9}.
When $P<P_{c1}$, the blue isobaric curve has two extremal points that divide it into three branches: the stable small BH branch, the unstable intermediate BH branch, and the stable large BH branch. When the pressure belongs to $P_{c1}<P<P_{c2}$, four extremal points appeared for the red isobaric curve, which divided it into five branches: the stable small BH branch, the unstable small BH branch, the stable intermediate BH branch, the unstable large BH branch, and the stable large BH branch. For $P_{c2}<P<P_{c3}$, two extremal points appear for the orange isobaric curve, which means the existence of three BH branches. When $P>P_{c3}$, the gray isobaric curve does not exhibit any extremal points, which implies there is only one BH branch.

\begin{figure} [H]
	\centering
	\subfigure[]{
		\includegraphics[width=0.4\textwidth]{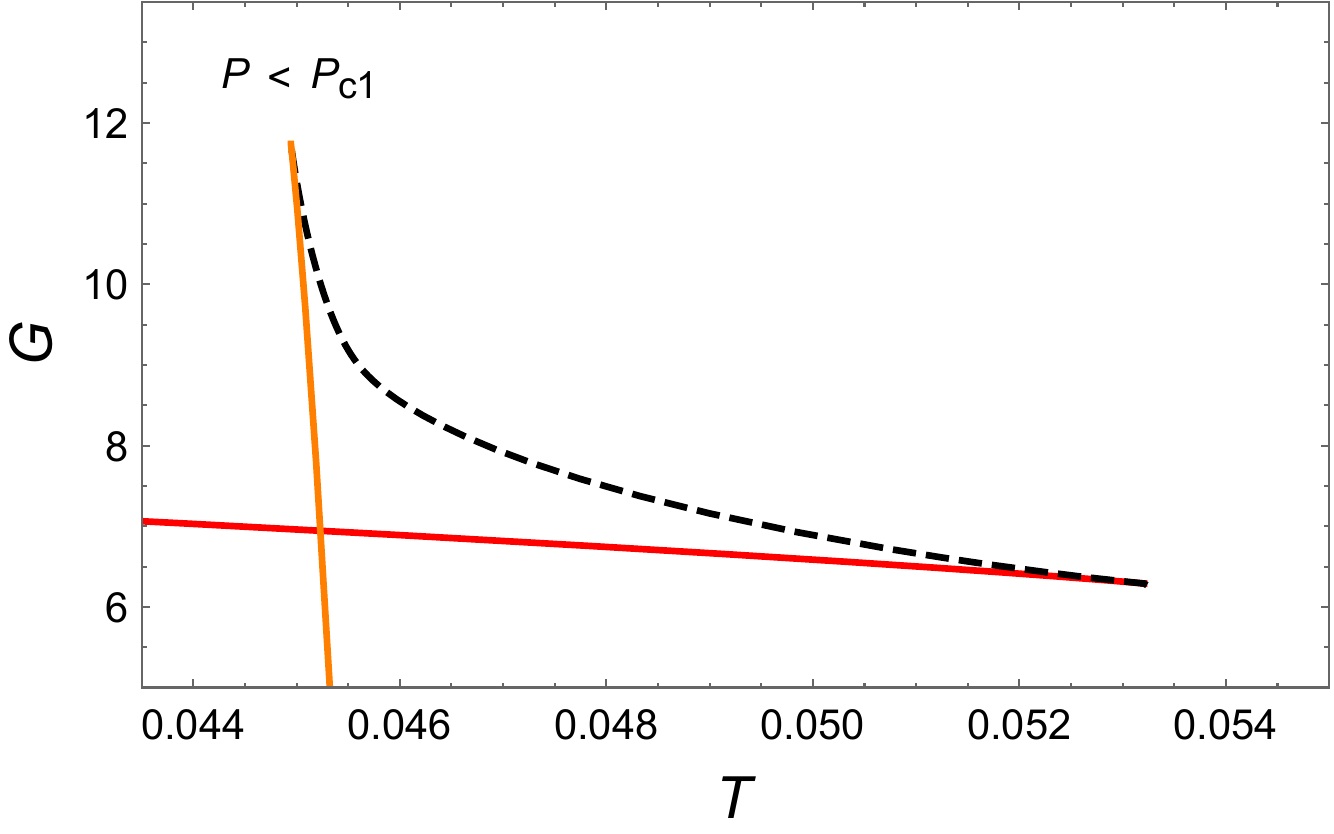}}
	\subfigure[]{
		\includegraphics[width=0.4\textwidth]{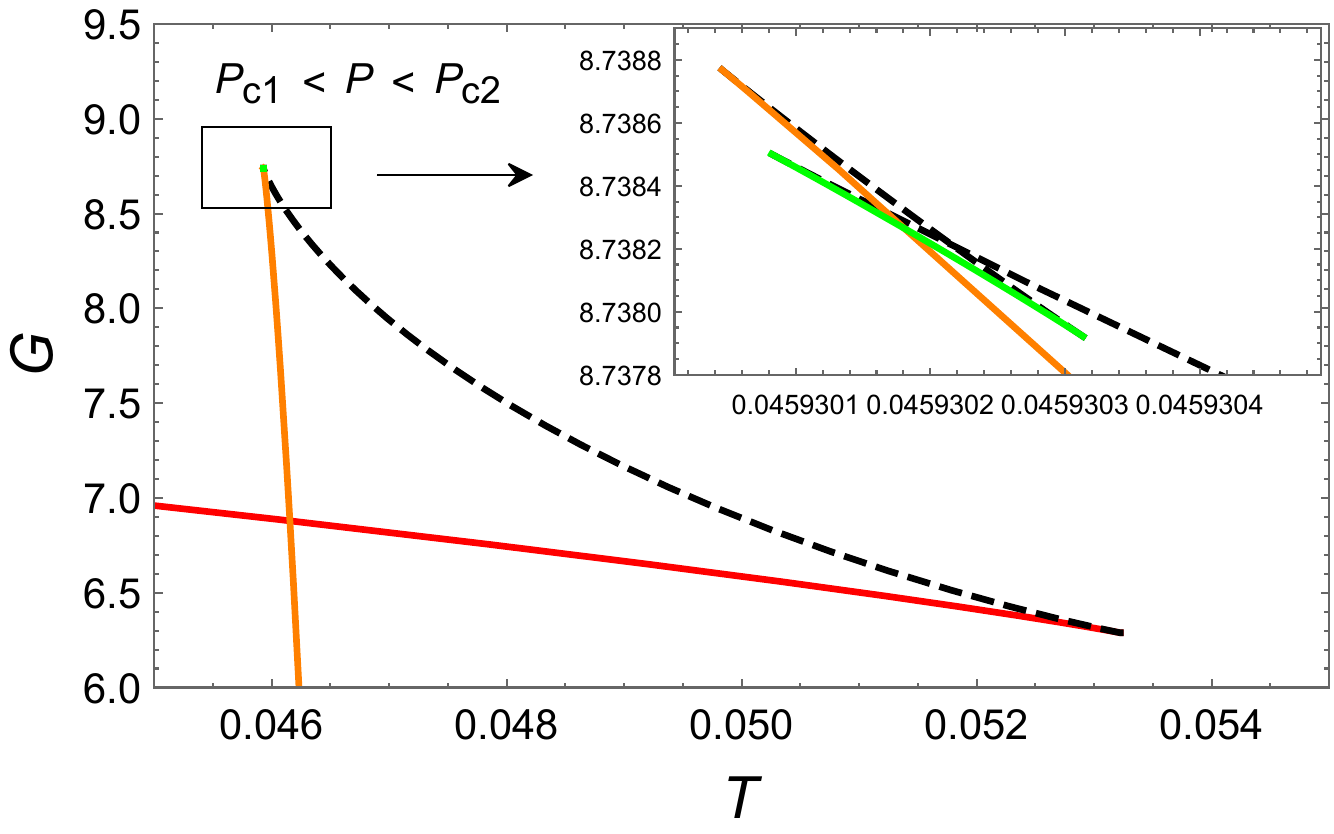}}
	\subfigure[]{
		\includegraphics[width=0.4\textwidth]{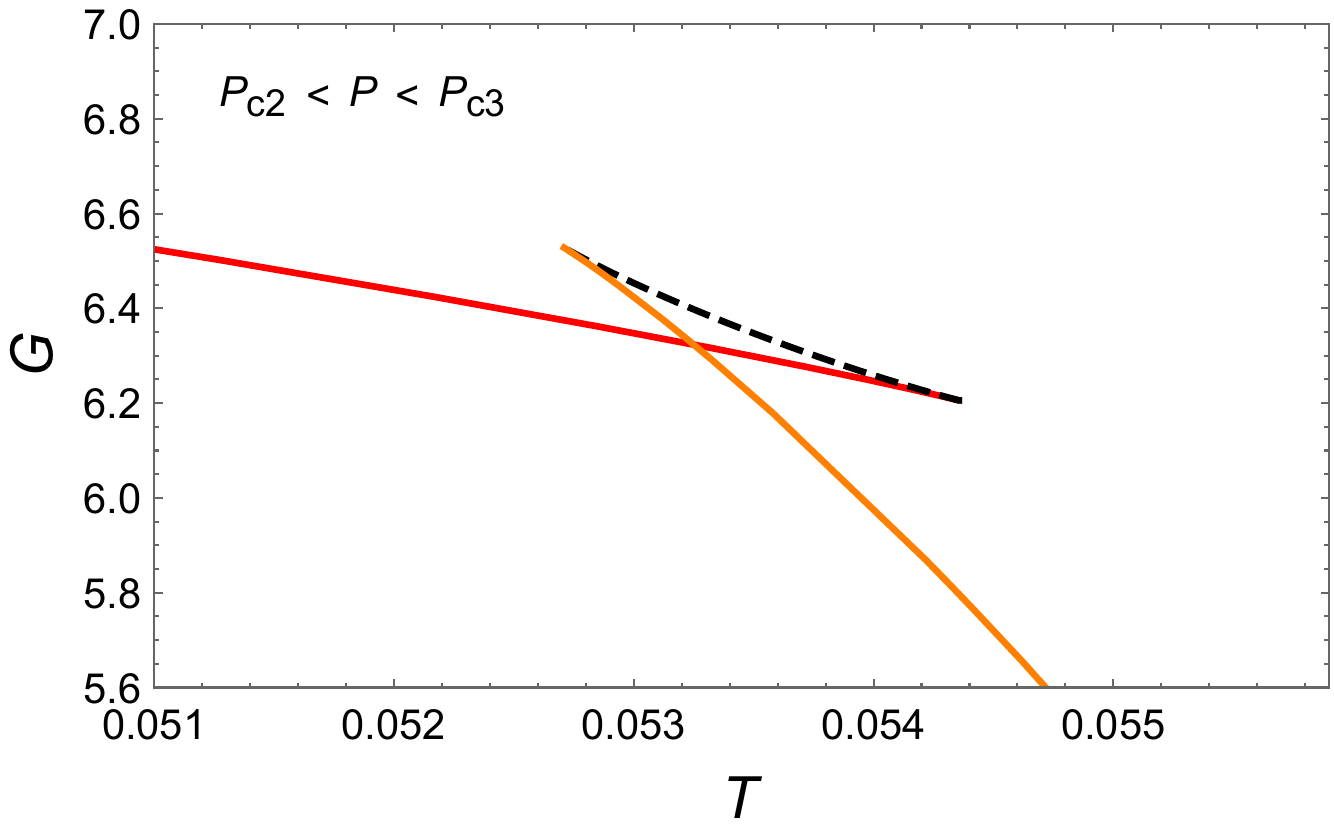}}
	\subfigure[]{
		\includegraphics[width=0.4\textwidth]{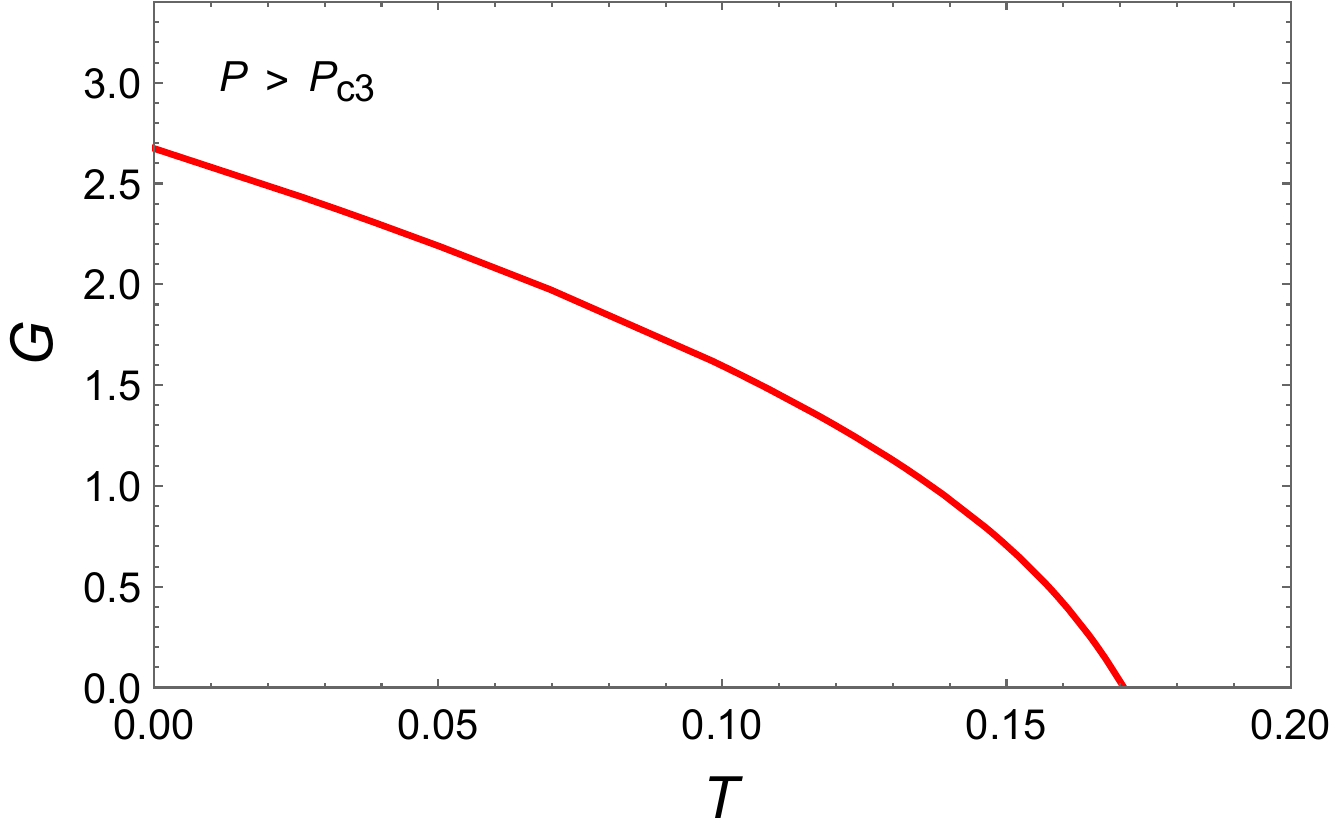}}
	\caption{\small Gibbs free energy $G$ vs. $T$. The red, green and orange curves represent small, intermediate and large BHs, respectively. (a) $P=0.003$. (b) $P=0.0033097$. (c) $P=0.025$. (d) $P=0.07$.}\label{fig10}
\end{figure}

\begin{figure} [H]
	\centering
	\subfigure{
		\includegraphics[width=0.38\textwidth]{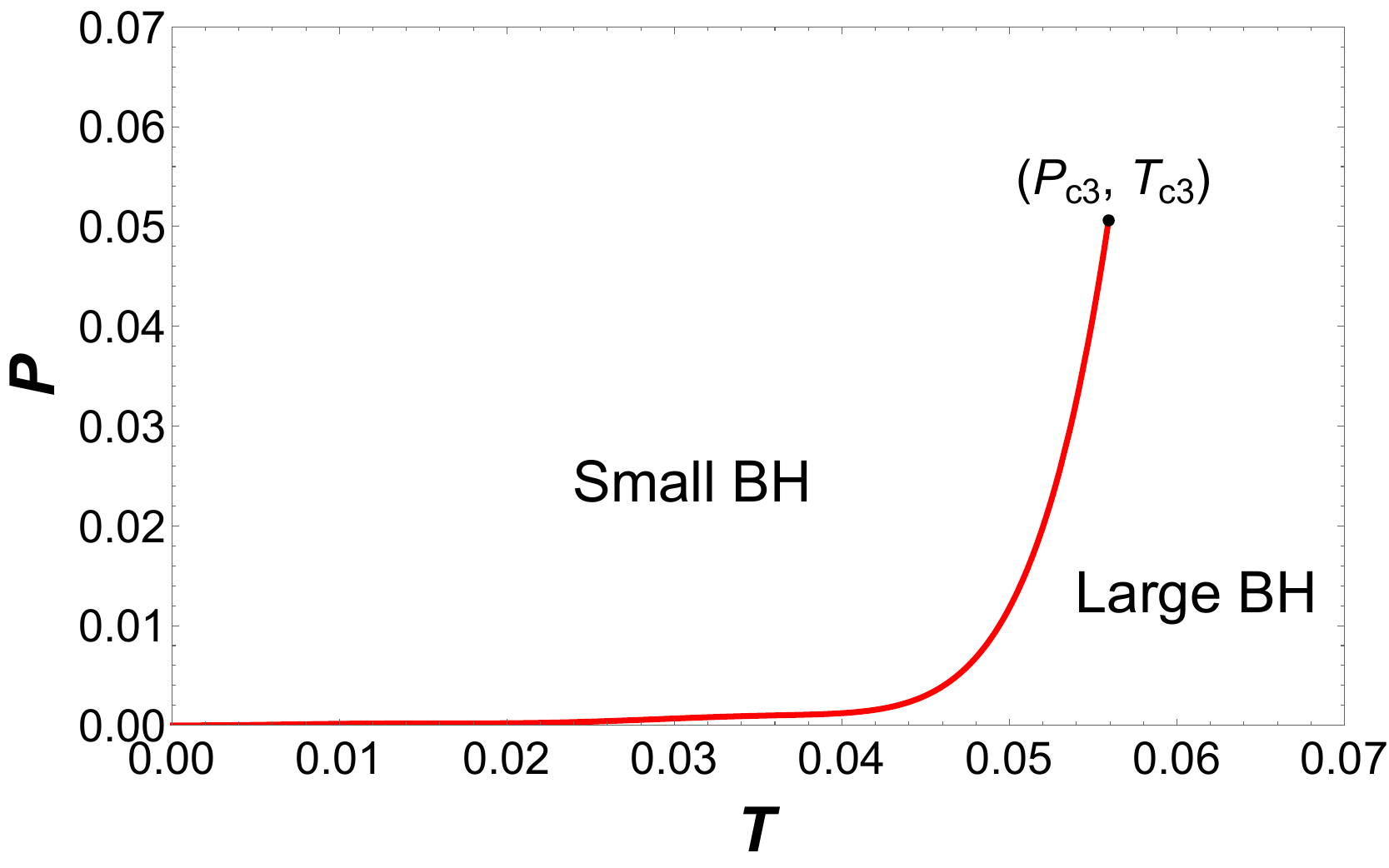}}
	\caption{\small Phase diagram for the 6-dimensional dyonic AdS BHs when $\alpha=1$ and $\beta=0.1$}\label{fig11}
\end{figure}

Next, we would like to study the behavior of Gibbs free energy. Therefore, the behavior of $G$ with respect to $T$ is plotted in Fig.\ref{fig10}. When $P<P_{c1}$, as shown in Fig.\ref{fig10}(a), the only one swallowtail indicates that the system undergoes a small/large BH phase transition. For $P_{c1}<P<P_{c2}$, as shown in Fig.\ref{fig10}(b), the two swallowtails seems indicates there are two first-order phase transitions in the system. However, the intermediate BH branch is suppressed by the BH branch with lower free energy due to its higher free energy, so it cannot participate in the phase transitions. As a result, only a small/large BH phase transition occurs in this pressure range. When the pressure increased upto $P_{c2}<P<P_{c3}$, as shown in Fig.\ref{fig10}(c), the only one swallowtail indicates that the system only undergoes a small/large BH phase transition. Once the pressure exceeds $P_{c3}$, the swallowtail behavior disappears, which leads to that no phase transition occurs in the system. In addition, we present the phase diagram in Fig.\ref{fig11}, where the coexistence curve terminates at the critical point ($P_{c3}, T_{c3}$). From it, one can obviously see that there has no triple point existed in current situation.
To sum up, when $\alpha=0.1$ and $1$, the BH system only undergoes a small/large BH phase transition, which is very different from the case of $\alpha=0.5$. Hence, it is true that the GB coupling constant $\alpha$ possess an important influence on the phase transition of the dyonic AdS BH.

\begin{figure} [H]
	\centering
	\subfigure[]{
		\includegraphics[width=0.38\textwidth]{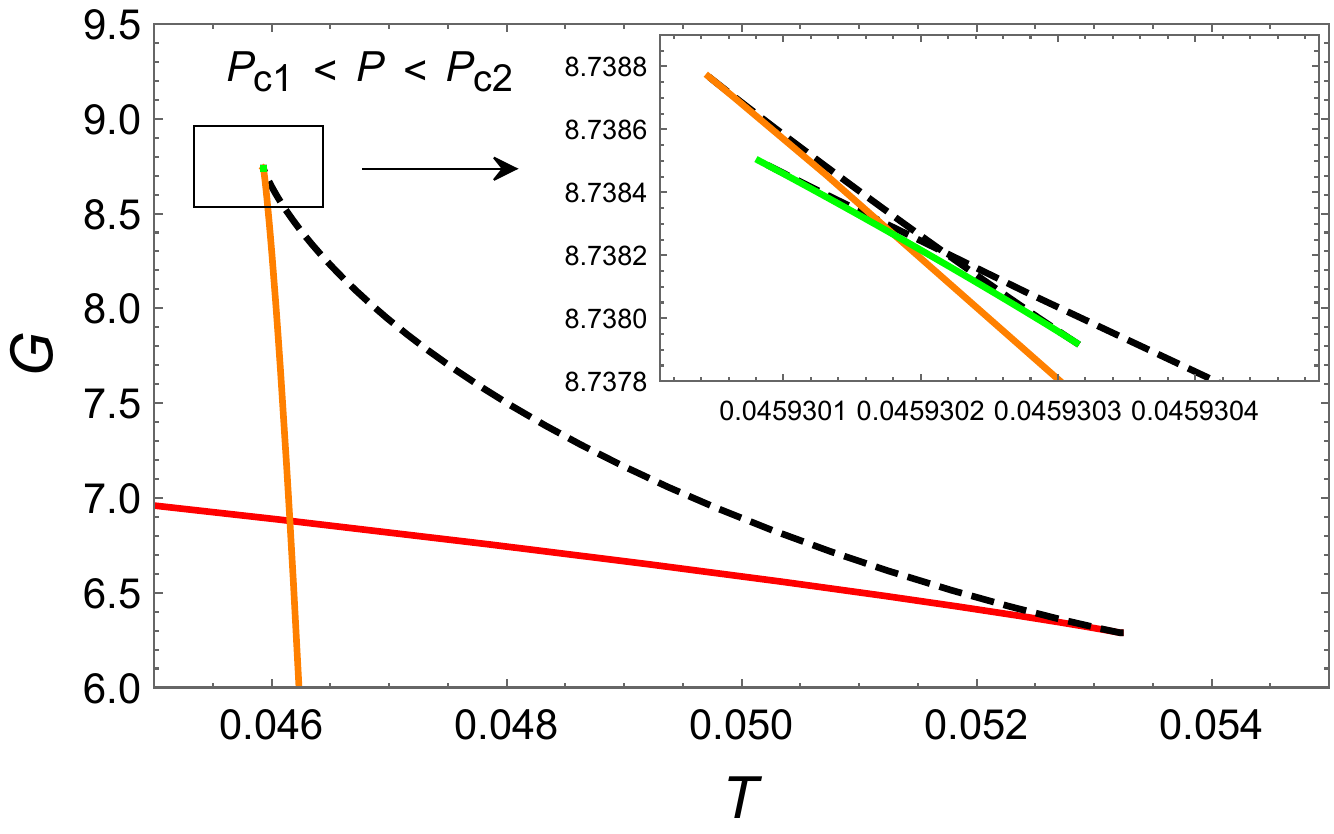}}
	\subfigure[]{
		\includegraphics[width=0.38\textwidth]{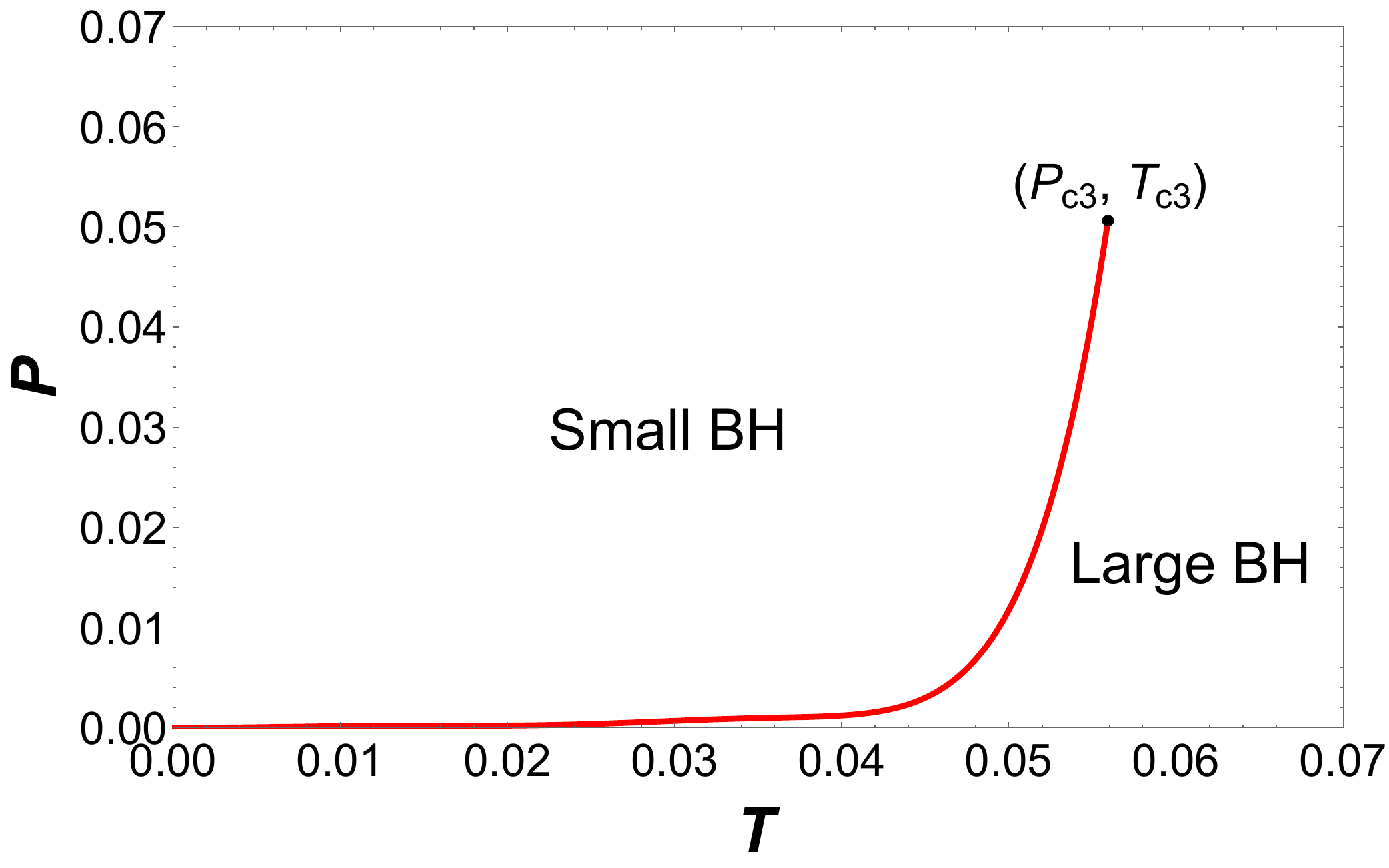}}
	\caption{\small (a) $G$ vs. $T$. (b) Phase diagram for the 6-dimensional dyonic AdS BH when $\alpha=1$ and $\beta=0.5$.}\label{fig12}
\end{figure}

In addition, we also studied the phase transitions for $\alpha=1$ and $\beta=0.5$, for which the $G-T$ diagram and $P-T$ phase diagram are illustrated in Fig.\ref{fig12}. In this case, we only present the $G-T$ diagram for the pressure of $P_{c1}<P=0.0033097<P_{c2}$, as it shown the fact that the intermediate BHs do not participate in the phase transition, resulting in only small/large BH phase transition occurs. Comparing Figs.\ref{fig10}, \ref{fig11} and \ref{fig12}, it can be observed that their results are almost identical, which also implies that the parameter $\beta$ has a little effect on the phase transition.
Additionally, when we further studied the phase transitions for $\alpha=10$ and $\alpha=50$, it is true that the corresponding results are similar to that of $\alpha=1$, where only the small/large BH phase transition occurs and $\beta$ also has a little effect on it. 

Finally, combined with the above results obtained in three examples ($\alpha=0.1$ and $\beta=0.1$; $\alpha=0.5$ and $\beta=0.1$; $\alpha=1$ and $\beta=0.1$), it can be concluded that in the context of 6-dimensional, the GB coupling constant $\alpha$ has a strong effect on the phase transition of BHs, while $\beta$ has a weaker one.

\subsection{For the $7$-dimensional case}\label{sec34}
\begin{figure} [H]
	\centering
	\subfigure[]{
		\includegraphics[width=0.3\textwidth]{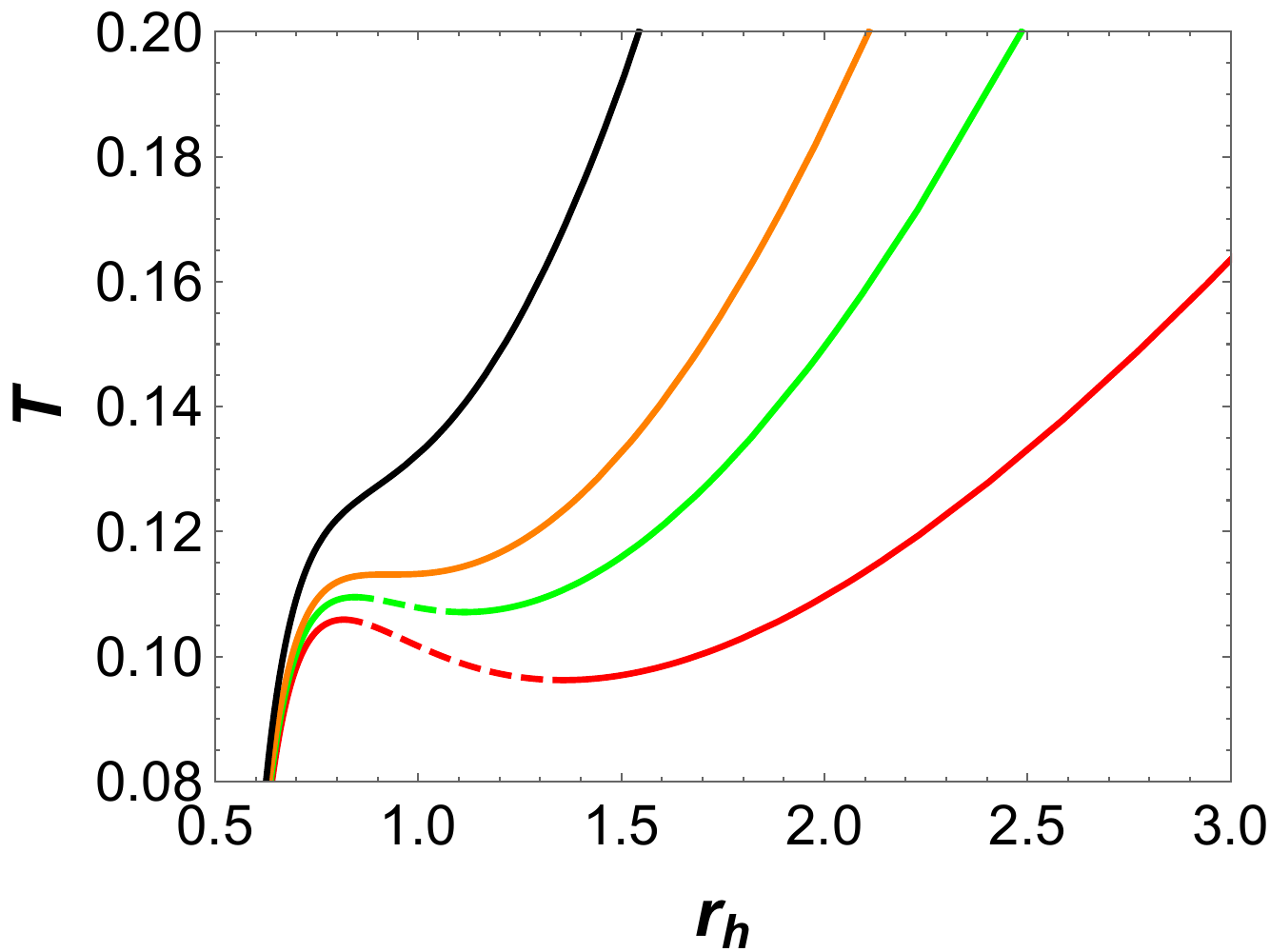}}
	\subfigure[]{
		\includegraphics[width=0.3\textwidth]{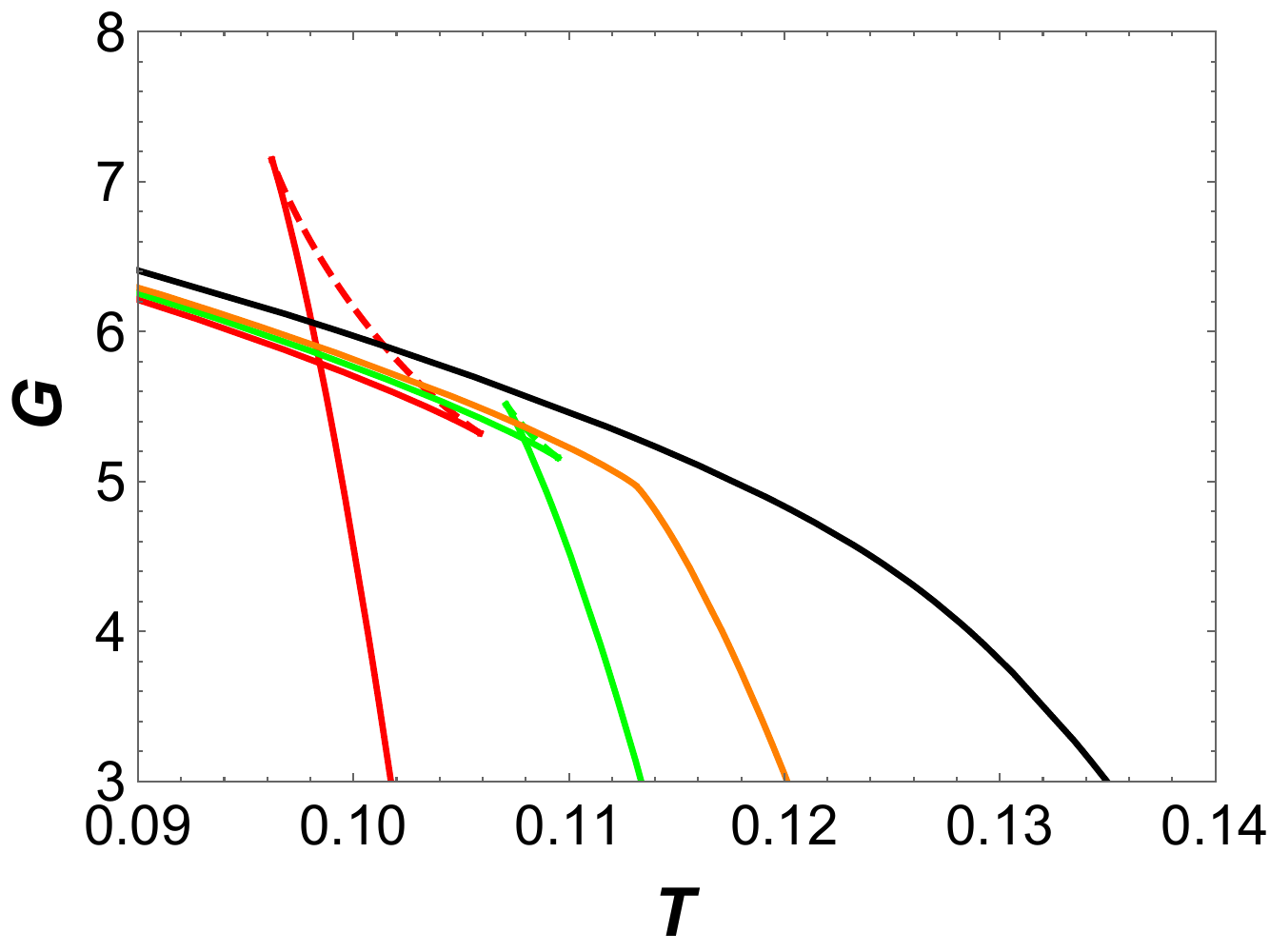}}
	\subfigure[]{
		\includegraphics[width=0.3\textwidth]{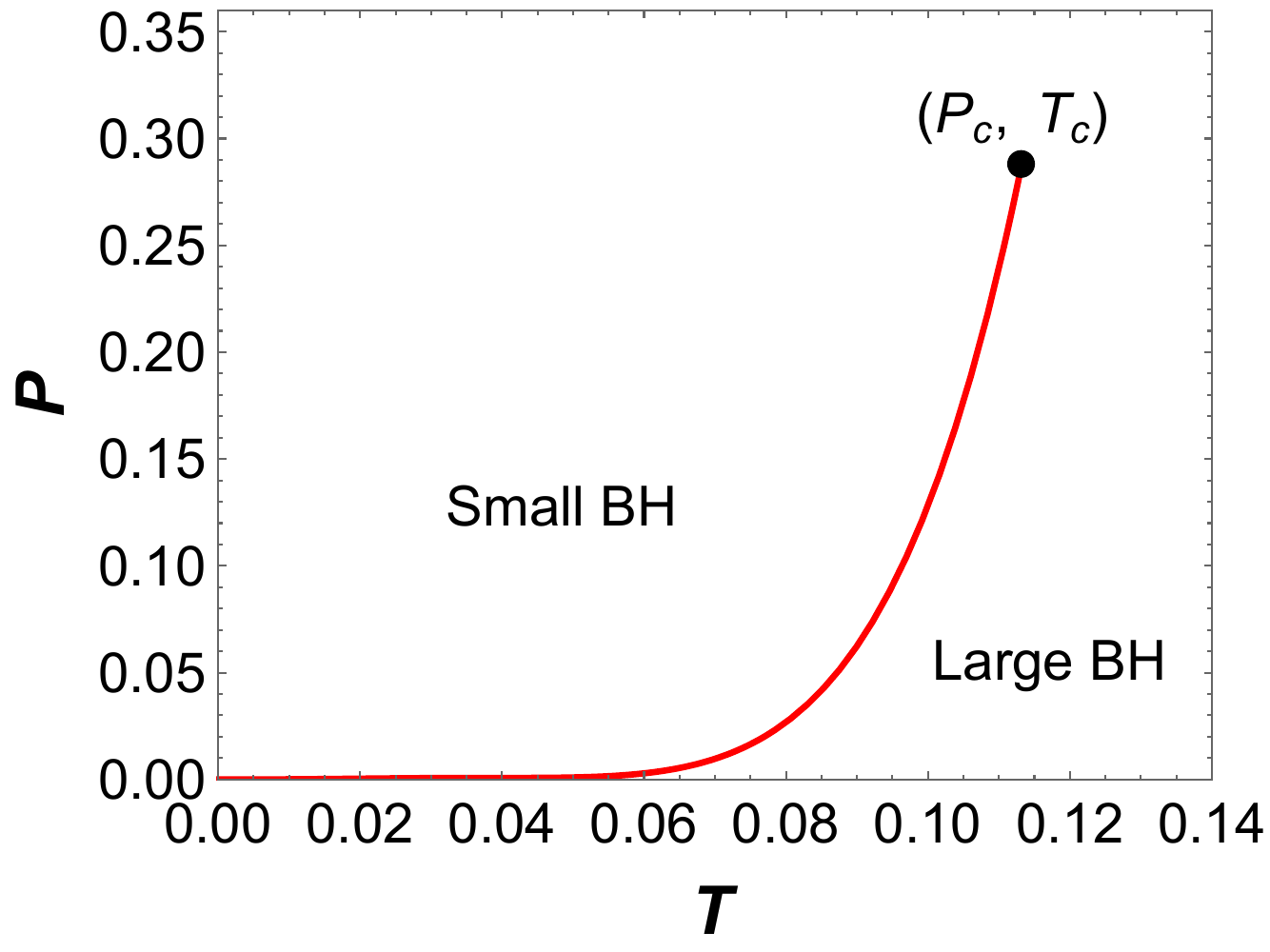}}
	\caption{\small (a) Temperature $T$ vs. $r_{h}$. (b) Gibbs free energy $G$ vs. $T$. The red, green, orange and black curves in (a) and (b) correspond to pressures $P=0.1$, $0.2$, $0.288091$ and $0.6$, respectively. (c) Phase diagram for the 7-dimensional dyonic AdS BH when $\alpha=0.5$ and $\beta=0.1$.
	}\label{fig13}
\end{figure}

In this subsection, we would like to study the phase transitions and phase diagrams of the 7-dimensional dyonic AdS BH. In a similar way, the critical points can be obtained, as shown in Tab.\ref{ta4}.
\begin{table}[H]
	\centering
	\captionsetup{font=footnotesize}
	{\caption {Critical points for different values of coupling constants $\alpha$ and $\beta$, when the dimension $D=7$.}
		{\footnotesize	\vspace{1mm}
			
			\begin{tabular} {ccc|ccc}
				\hline 
				{$\alpha$}  &{$\beta$}     &{$(T_{c},P_{c})$}  &{$\alpha$}  &{$\beta$}     &{$(T_{c},P_{c})$}   \\ 
				\hline
				{0.01}   &{0.01}  &{(0.3468615086, 0.1399764527)}  &{0.1}    &{0.01}   &{(0.1536583170, 0.04142045303)} \\
				{0.01}   &{0.1}   &{(0.3468615086, 0.1399764527)}  &{0.1}    &{0.1}    &{(0.1536583170, 0.04142045303)}  \\
				{0.01}   &{0.5}   &{(0.3468615086, 0.1399764527)}  &{0.1}    &{0.5}    &{(0.1536583170, 0.04142045303)}  \\
				{0.01}   &{1}     &{(0.3468615086, 0.1399764527)}  &{0.1}    &{1}      &{(0.1536583170, 0.04142045303)}   \\
				{0.01}   &{10}    &{(0.3468615086, 0.1399764527)}  &{0.1}    &{10}     &{(0.1536583170, 0.04142045303)}   \\
				{0.01}   &{50}    &{(0.3468615086, 0.1399764527)}  &{0.1}    &{50}     &{(0.1536583170, 0.04142045303)}   \\
				\hline
				{0.5}	 &{0.01}  &{(0.1131229239, 0.2880911532)}  &{1}      &{0.01}   &{/}  \\
				{0.5}	 &{0.1}   &{(0.1131229239, 0.2880911532)}  &{1}      &{0.1}    &{/}  \\
				{0.5}	 &{0.5}   &{(0.1131229239, 0.2880911532)}  &{1}      &{0.5}    &{/}  \\
				{0.5}	 &{1}     &{(0.1131229239, 0.2880911532)}  &{1}      &{1}      &{/}  \\
				{0.5}	 &{10}    &{(0.1131229239, 0.2880911532)}  &{1}      &{10}     &{/}   \\
				{0.5}    &{50}    &{(0.1131229239, 0.2880911532)}  &{1}      &{50}     &{/}      \\
				\hline
				{10}	 &{0.01}  &{/}                             &{50}     &{0.01}   &{/}  \\
				{10}	 &{0.1}   &{/}                             &{50}     &{0.1}    &{/} \\
				{10}	 &{1}     &{/}                             &{50}     &{1}      &{/}  \\
				{10}     &{50}    &{/}                             &{50}     &{50}     &{/}      \\
				\hline 
			\end{tabular}\label{ta4}}		
}\end{table}
In Tab.\ref{ta4}, when $\alpha$ fixed and $\beta$ varied, the critical points remain constant. However, when $\alpha\geqslant1$, there is no critical point. It is no doubt that the BH system exhibits a single phase structure in dimension $D=7$. In the case, i.e., $\alpha=0.5$ and $\beta=0.1$, the critical point is given as
\begin{align}
	T_{c}=0.113123,P_{c}=0.288091.
\end{align}
The behavior of temperature $T$ with respect to $r_{h}$ and Gibbs free energy $G$ with respect to $T$ are plotted in Figs.\ref{fig13}(a) and \ref{fig13}(b), respectively. From the Fig.\ref{fig13}, it can be see that their behaviors are similar to that in dimension $D=4$ and $5$. When $P<P_{c}$, the red and green isobaric curves in the $T-r_{h}$ diagram are divided into three branches by two extremal points, and only one swallowtails appear in the $G-T$ diagram for each curve. 
For $P=P_{c}$, the system undergoes a second-order phase transition. When $P>P_{c}$, the black isobaric curve in the $T-r_{h}$ diagram become monotonic functions of $r_{h}$. These results indicate that a small/large BH phase transition occurs in the BH system when $D=7$. All of these results are similar with that of $D=4,5$, and the corresponding phase diagram is shown in Fig.\ref{fig13}(c).

\section{Critical exponents}\label{sec4}
In this section, we would like to calculate the critical exponents near the critical points. It is well known that critical exponents provide an effective way to characterize the behavior of physical quantities in the vicinity of the critical point. In general, these exponents are universal, and the details of the physical system would not effect it.

For simplicity, we define a few reduced parameters,
\begin{align}
	t=\frac{T}{T_{c}}-1=\tau-1,\label{reduced-temperature}
\end{align}
where $\tau=\frac{T}{T_{c}}$ is the reduced thermodynamic temperature,
\begin{align}
	\omega=\frac{V}{V_{c}}-1=\nu-1,\label{reduced-volume}
\end{align}
where $\nu=\frac{V}{V_{c}}$ is the reduced thermodynamic volume, and 
\begin{align}
	p=\frac{P}{P_{c}}\label{reduced-pressure}
\end{align}
is the reduced thermodynamic pressure.

Next, let us review the definitions of the critical exponents $\alpha_{1}$, $\beta_{1}$, $\gamma$, and $\delta$ near the critical point, which can be expressed as follows\cite{Kubiznak:2012wp}:
\\(1) Exponent $\alpha_{1}$ determines the behavior of the specific heat at constant volume,
\begin{align}
	C_{V}=T\frac{\partial S}{\partial T}|_{V}\varpropto|t|^{-\alpha_{1}}.
\end{align}
(2) Exponent $\beta_{1}$ describes the behavior of the order parameter $\eta=V_{l}-V_{s}$ (the difference between the volumes of the coexisting large and small BHs) on a given isotherm
\begin{align}
	\eta=V_{l}-V_{s}\varpropto|t|^{\beta_{1}}.
\end{align}
(3) Exponent $\gamma$ governs the behavior of the isothermal compressibility $\kappa _T$,
\begin{align}
	\kappa _T=-\frac{1}{V}\frac{\partial V}{\partial P}|_{T}\varpropto|t|^{-\gamma}.
\end{align}
(4) Exponent $\delta$ reflected the following behavior on the critical isotherm $T=T_c$ is
\begin{align}
	|P-P_{c}|\varpropto|V-V_{c}|^{\delta}.
\end{align}

We will now solve these critical exponents one by one. To calculate $C_{V}$, we start from the definition of Helmholtz free energy $F(T,V)=G-PV$, and give the expression for Helmholtz free energy as
\begin{align}
	F(T,V)&= \frac{\pi^{\frac{D-3}{2}}r_{h}^{D-3}(D-2-4\pi r_{h}T)}{4(D-3)\Gamma(\frac{D-3}{2})}+ \frac{\pi^{\frac{D-3}{2}}r_{h}^{3-D}(Q_{m}^{2}+Q_{e}^{2}\mathcal{E}_{r_h})}{8(D-3)^{2}\Gamma(\frac{D-3}{2})}\nonumber \\
	&\quad+ \frac{(D-2)\pi^{\frac{D-3}{2}}r_{h}^{D-5}(D-4-8\pi r_{h}T)\alpha}{4\Gamma(\frac{D-3}{2})}. 
\end{align}
So, the entropy is
\begin{align}
S(T,V)=-\left(  \frac{\partial F}{\partial T}\right)  _{V}=\frac{\pi^{\frac{D-1}{2}} r_{h}^{D-2}[1+2(D-3)(D-2)\alpha r_{h}^{-2} ] }{2\Gamma \left( \frac{D-1}{2}\right)},
\end{align}
which is consistent with Eq.(\ref{entropy}). Since $S$ is independent of $T$, we have $C_{V}=0$, therefore the exponent $\alpha_{1}=0$.
To compute the exponents $\beta_{1}$, $\gamma$ and $\delta$, we substitute the reduced parameters introduced in Eqs.(\ref{reduced-temperature}), (\ref{reduced-volume}) and (\ref{reduced-pressure}) into Eq.(\ref{pressure}) to derive the corresponding equation of state. In the vicinity of the critical point, the reduced pressure can be expressed as
\begin{align}
	p=A_{0}+A_{1}\omega+A_{2}\omega^{2}+A_{3}\omega^{3}+B_{0}t+B_{1}t\omega+\mathcal{O}(t\omega^{2}, \omega^{4}).\label{reduced-pressure2}
\end{align}
The values of the expanded coefficients in Eq.(\ref{reduced-pressure2}) are calculated for various dimensions $D$, GB coupling constant $\alpha$ and coupling constant $\beta$, and the corresponding results are listed in Tab.\ref{ta5}. 

\begin{table}[H]
	\centering
	\captionsetup{font=footnotesize}
	{\caption {The values of the expanded coefficients in Eq.(\ref{reduced-pressure2}) for various $D$, $\alpha$ and $\beta$.}
		{\footnotesize	\vspace{1.5mm}
			
			\begin{tabular} {ccccccc}
				\hline 
				{$D$}  &{$\alpha$}  &{$\beta$}  &{$A_{0}$}  &{$A_{3}$}     &{$B_{0}$} &{$B_{1}$} \\ 
				\hline
				{4}   &{0.5}     &{0.1}   &{1}   &{-0.0493816}    &{2.66660}    &{-0.88887}    \\
				{5}   &{0.5}     &{0.1}   &{1}   &{-0.0184683}    &{3.70884}    &{-1.33622}    \\
				{6}   &{0.1}     &{0.1}   &{1}   &{-0.0118259}    &{4.10056}    &{-1.29675}    \\
				{6}   &{0.5}     &{0.1}   &{1}   &{-0.0190869}    &{24.6504}    &{-12.1297}     \\
				{6}   &{0.5}     &{0.1}   &{1}   &{-0.0031754}    &{6.63092}    &{-2.52051}     \\
				{6}   &{1}       &{0.1}   &{1}   &{-0.1408537}    &{14.0260}    &{-7.97729}     \\
				{6}   &{1}       &{0.1}   &{1}   &{-0.0013238}    &{7.38146}    &{-2.89173}     \\
				{7}   &{0.5}     &{0.1}   &{1}   &{-0.0903225}    &{7.85837}    &{-3.75320}     \\	
				\hline 
			\end{tabular}\label{ta5}}		
}\end{table}
As shown in Tab.\ref{ta5}, coefficient $A_{0}=1$, coefficient $B_{0}$ is positive, while coefficients $A_{3}$ and $B_{1}$ are negative. It is worth noting that coefficients $A_{1}$ and $A_{2}$ are absent in our BH system, hence they are not listed in the table. Therefore, we can reexpress the reduced pressure as
\begin{align}
	p=1+A_{3}\omega^{3}+B_{0}t+B_{1}t\omega+\mathcal{O}(t\omega^{2}, \omega^{4}).\label{reduced-pressure3}
\end{align}
Under the condition of fixing $t<0$, differentiating the reduced pressure in Eq.(\ref{reduced-pressure3}) yields
\begin{align}
	dp=(3A_{3}\omega^{2}+B_{1}t)d\omega.
\end{align}
In addition, the volumes of the coexisting small and large BHs are denoted by $\omega_{s}$ and $\omega_{l}$, respectively. The application of Maxwell equal area law gives 
\begin{align}
	\int_{\omega_{s}}^{\omega_{l}}\omega(3A_{3}\omega^{2}+B_{1}t)d\omega=0.\label{eq49}
\end{align}
 Moreover, the coexisting small and large BHs satisfy the equation of state, which is expressed as
 \begin{align}
 	p=1+A_{3}\omega_{s}^{3}+B_{0}t+B_{1}t\omega_{s}=1+A_{3}\omega_{l}^{3}+B_{0}t+B_{1}t\omega_{l}.\label{eq50}
 \end{align}
By solving Eqs.(\ref{eq49}) and (\ref{eq50}), we get $\omega_{s}=-\omega_{l}=-\sqrt{\frac{B_{1}}{A_{3}}}\sqrt{-t}$. Therefore, the order parameter $\eta$ satisfies 
\begin{align}
	\eta=V_{c}(\omega_{l}-\omega_{s})=2V_{c}\sqrt{\frac{B_{1}}{A_{3}}}\sqrt{-t},\label{eq51}
\end{align}
which indicates that the exponent $\beta_{1}=\frac{1}{2}$.
To compute the exponent $\gamma$, we differentiate Eq.(\ref{reduced-pressure3}),  and then get
\begin{align}
	\frac{\partial V}{\partial P}|_{T}=\frac{1}{B_{1}}\frac{V_{c}}{P_{c}}\frac{1}{t}+\mathcal{O}(\omega).\label{eq52}
\end{align}
 Thus, 
 \begin{align}
 	\kappa_{T}=-\frac{1}{V}\frac{\partial V}{\partial P}|_{T}\propto-\frac{1}{B_{1}}\frac{V_{c}}{P_{c}}\frac{1}{t},
 \end{align}
  which implies that the exponent $\gamma=1$. Finally, by setting the reduced temperature $t=0$ in Eq.(\ref{reduced-pressure3}), we obtain 
 \begin{align}
 	p-1=A_{3}\omega^{3},
 \end{align} 
 which indicates that the exponent $\delta=3$. It can be easily verified that these critical exponents satisfy the following scaling laws of thermodynamics,
 \begin{align}
 	\alpha_{1}+2\beta_{1}+\gamma=2, \
 	\alpha_{1}+\beta_{1}(1+\delta)=2,  \nonumber \\
	\gamma(1+\delta)=(2-\alpha_{1})(\delta-1),  \
	\beta_{1}(\delta-1)=\gamma.
\end{align}
In summary, after calculating the four critical exponents of our BH system, namely
 \begin{align}
 	\alpha_{1}=0, \
 	 \beta_{1}=\frac{1}{2}, \ 
 	 \gamma=1,  \
 	 \delta=3,
 \end{align} 
  we conclude that our BH system has the same critical exponents which have also be obtained in both the four-dimensional charged RN AdS BH system\cite{Kubiznak:2012wp} and the four-dimensional dyonic AdS BH system\cite{Li:2022vcd}. In fact, these critical exponents possess identical values which is coincide with the mean field theory. This property is very important which further greatly support our results obtained in this paper.
  
\section{Conclusion and discussion}\label{sec5}
In this paper, we have investigated the thermodynamics and phase transitions of the dyonic AdS BHs with quasitopological electrodynamics in EGB gravity, where the negative cosmological constant $\Lambda$ is regarded as the thermodynamic pressure of BH and its conjugate quantity is introduced as the thermodynamic volume of BH. By treating the GB coupling constant $\alpha$ and coupling constant $\beta$ as novel thermodynamic variables, we checked the first law of thermodynamics and derived the generalized Smarr relation in this BH. Then, we studied the phase transitions by analyzing the characteristics of temperature and Gibbs free energy for the dyonic AdS BHs.
Specifically, we studied the phase transitions and phase diagrams for different dimensional cases, i.e., 4, 5, 6 and 7.

In the 4-dimensional case, we observed a typical small/large BH phase transition, which is similar to the vdW liquid/gas phase transition. Notably, we find that the coupling constant $\beta$ has a stronger effect on the critical points in contrast to $\alpha$.
The BH phase transition for the 5-dimensional case is similar to the 4-dimensional one, but the effect of $\alpha$ on the critical points is larger than that of $\beta$, rather than a smaller one.
Interestingly, we also discovered the typical small/large BH phase transition for the 7-dimensional case, where $\alpha$ has a greater impact on the critical points by comparing with $\beta$.
However, the critical points in the 6-dimensional case are more interesting than those in the 4, 5 and 7-dimensional cases. Specifically, one can see that three critical points are obtained in the parameter region where $\alpha\geqslant0.5$, which implies that the system exhibits some richer phase behaviors. 
It also can be see that $\alpha$ has a significant effect on the values of the critical points, while the effect of $\beta$ is weaker.
For convenience, by fixing $\beta=0.1$, we taken $\alpha=0.1$, $0.5$ and $1$ as three typical examples to further study the phase transitions and phase diagrams of the dyonic AdS BHs.
For $\alpha=0.1$, we find there is a typical small/large BH phase transition in this case. 
And, it shows for $\alpha=0.5$ that the system undergoes the small/intermediate/large BH phase transitions, rather than the small/large BH phase transition. Also, the triple point where all three BH phases coexisted is obtained, i.e., ($P_t=0.006469822191382416, T_t=0.06470845191069545$).
In the case $\alpha=1$, although we find three critical points in this system, there is only the small/large BH phase transition, thereby the triple point no longer exists.
When $\alpha>1$, i.e., $\alpha=10$ and $50$, the results show that the BH phase transition is similar to that obtained in the case ($\alpha=1$).
Meanwhile, it is true in 6-dimensional space-time that the coupling constant with respect to quasitopological electrodynamics $\beta$ almost has no influence on the critical points, as well as phase transitions.
This implies the effect of the parameter $\alpha$ on the BH phase transition is larger than that of $\beta$ in the 6-dimensional case for the dyonic AdS BHs. 
Combined with above facts, we conclude that the GB coupling constant $\alpha$ and space-time dimension $D$ play as two key roles on the study of the phase transitions and phase diagrams of the dyonic AdS BHs with quasitopological electrodynamics in EGB gravity, for instance, the phase structure of the dyonic AdS BHs in the case ($D=6$ and $\alpha=0.5$) is much more complex and interesting than that in other dimensions and choices of $\alpha$.

Furthermore, after calculating the critical exponents near the critical points, we obtain $\alpha_{1}=0$, $\beta_{1}=\frac{1}{2}$, $\gamma=1$, and $\delta=3$. 
Clearly, the critical exponents share the same values as mean field theory, and which is also full consistence with that of other BH systems. 
Finally, we conclude that our results can provide some deep insights into the intriguing thermodynamic properties of the dyonic AdS BHs with quasitopological electromagnetism in EGB gravity.

In addition, the holographic duality also provides a valuable approach for the study of the thermodynamics of AdS BHs\cite{Ahmed:2023snm,Ahmed:2023dnh}.
Therefore, it is meaningful to further study and test the thermodynamics of the dyonic AdS BHs in the framework of AdS/CFT correspondence, which may further reveal more interesting thermodynamic properties. And, this will be included in our future work.

\vspace{11pt}

\noindent {\bf Acknowledgments}

This work is supported by the National Natural Science Foundation of China (Grant No. 11903025), and by the starting fund of China West Normal University (Grant No.18Q062), and by the Sichuan Youth Science and Technology Innovation Research Team (21CXTD0038)and by the Natural Science Foundation of SiChuan Province(2022NSFSC1833).

\end{document}